\documentclass[acmtog]{acmart}

\title{Seamless Parametrization in Penner Coordinates}
\setcitestyle{square}

\usepackage{acmart-taps}

\usepackage[utf8]{inputenc}

\usepackage{natbib}
\usepackage{subcaption}

\usepackage{xcolor}
\usepackage{ifthen}
\usepackage{siunitx}
\usepackage[capitalize,noabbrev]{cleveref}

\usepackage{multirow}
\usepackage{makecell} 
\usepackage{pgfplots}
\pgfplotsset{compat=1.18}
\usepackage{wrapfig}
\usepackage{csquotes}
\usepackage[ruled,vlined,noend,linesnumbered]{algorithm2e}
\usepackage{setspace}
\usepackage{graphicx}
\usepackage{overpic}
\usepackage[]{mdframed}

\definecolor{DenisColor}{rgb}{0.15, 0.38, 0.68}
\definecolor{RyanColor}{rgb}{0.88, 0.66, 0.43}

\newcommand{\RC}[1]{}
\newcommand{\DZ}[1]{}

\aptLtoXcmd{}{%
}

\aptLtoXcmd{}{%
}

\aptLtoXcmd{}{}

\usepackage{acronym}
\newacro{IP}{incremental potential}
\newacro{IPC}{Incremental Potential Contact}
\newacro{CIPC}[C-IPC]{Codimensional \ac{IPC}}
\newacro{NURBS}{non-uniform rational B-spline}
\newacro{IGA}{isogeometric analysis}
\newacro{DOF}{degrees of freedom}
\newacro{FE}{finite element}
\newacro{FEM}{finite element method}
\newacro{CCD}{continuous collision detection}
\newacro{LCP}{linear complementarity problem}
\newacro{LBFGS}[L-BFGS]{Limited-memory BFGS}
\newacro{HPC}{high performance computing}
\newacro{HO}{high-order}

\graphicspath{{./figs/}{./figs/sketches}}

\newcommand{\bR}{\mathbb{R}}

\newcommand{\cP}{\mathcal{P}}

\newcommand{\tl}{{\tilde{\lambda}}}
\newcommand{\tM}{{\tilde{M}}}
\newcommand{\scp}[1]{{\scshape{#1}}}
\newcommand{\hTh}{\hat\Theta}

\newcommand{\del}{\mathrm{Del}}

\newcommand{\DF}{\nabla F}

\setcopyright{acmlicensed}
\acmJournal{TOG}
\acmYear{2024} \acmVolume{43} \acmNumber{4} \acmArticle{61} \acmMonth{7}\acmDOI{10.1145/3658202}



\author{Ryan Capouellez}
\email{rjc8237@nyu.edu}
\affiliation{%
  \institution{New York University}
  \country{USA}
}

\author{Denis Zorin}
\email{dzorin@cs.nyu.edu}
\affiliation{%
  \institution{New York University}
  \country{USA}
}

\AtBeginMaketitle{%
\let\oldAtBeginDocument\AtBeginDocument%
\renewcommand\AtBeginDocument[1]{#1}
  
\let\AtBeginDocument\oldAtBeginDocument%
}

\begin{document}

\begin{abstract}

We introduce a conceptually simple and efficient algorithm for seamless parametrization,  a key element in constructing quad layouts and texture charts on surfaces. More specifically, we consider the construction of parametrizations with prescribed \emph{holonomy signatures} i.e., a set of angles at singularities, and rotations along homology loops, preserving which is essential for constructing parametrizations following an input field, as well as for user control of the parametrization structure. 
Our algorithm performs exceptionally well on a large dataset based on Thingi10k \cite{Thingi10K}, (16156 meshes) as well as on a challenging smaller dataset of \cite{Myles:2014}, converging, on average, in 9 iterations. 
Although the algorithm lacks a formal mathematical guarantee, presented empirical evidence and the connections between convex optimization and closely related algorithms, suggest that a similar formulation can be found for this algorithm in the future.  
\end{abstract}

\begin{teaserfigure}
\includegraphics[width=\textwidth]{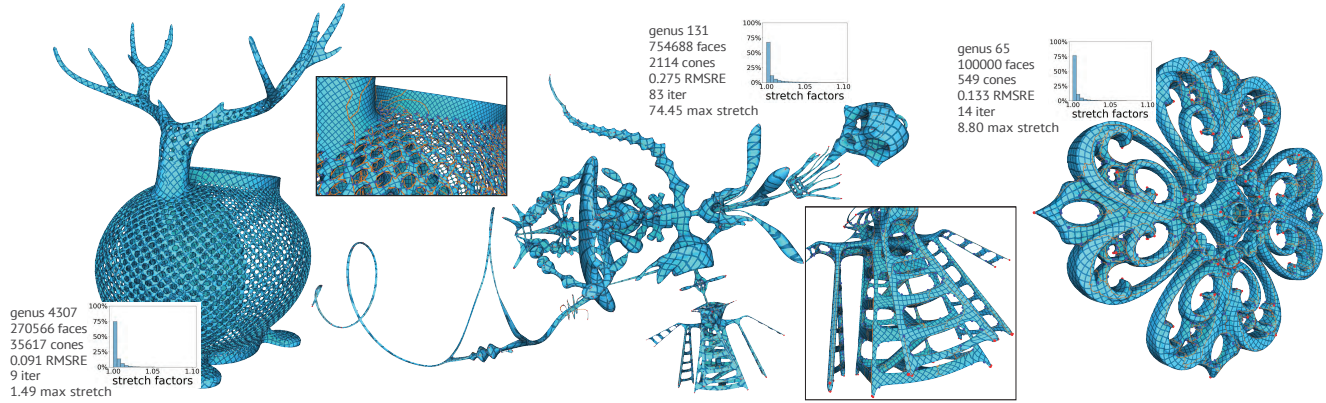}
\label{fig:teaser}
\caption{Our algorithm is demonstrated to work reliably on a broad range of challenging models, with high geometric complexity and high genus. Three examples shown above: 
the highest-genus model from Thingi10k dataset \protect\citep{Thingi10K}, "yeahright" model courtesy of Keenan Crane, and the Filigree model from the dataset of \citet{Myles:2014}.}
\end{teaserfigure}

\maketitle

\acresetall 

\section{Introduction}
\label{sec:intro}

Seamless parametrization is an important starting point for a widely used family of algorithms for constructing quad layouts. It can be applied in any context that requires an atlas of texture charts on the surface with complete freedom for choosing the cuts with no or minimal transition artifacts.  Informally, a global parametrization i.e., a locally injective map from a cut surface to the plane, is seamless, if the parametric lines along $u$ and $v$ directions continue smoothly across the cut, and the parametric lengths of cut edges match. 

A geometrically natural way to define a seamless parametrization is as a metric on the mesh (i.e., an assignment of \emph{lengths} to edges) with angle constraints.  A map to the plane corresponds to the angle constraint that at almost all vertices the sum of angles is $2\pi$. The vertices where the sum is not $2\pi$ (\emph{cones}) are necessary due to topological reasons for all surfaces of genus $g \neq 1$.  \emph{Seamless} parametrizations require additional conditions on sums of angles at cones and more generally along all dual loops: these need to be multiples of $\pi/2$ if added with appropriate signs (Section~\ref{sec:method}). 
There is a basis of $n_c + 2g$ loops, $n_c$ loops around cone vertices and $2g$ homology basis loops, that completely determines the angles on all dual loops.

Following \citet{Campen:2017:SimilarityMaps}, we refer to this collection of $n_c + 2g$
angles of the form $k_i \pi/2$ as \emph{holonomy signature}.
One can think about this as defining the coarse topology of a seamless parameterization: e.g., if a parameterization is used to obtain a quadrangulation by tracing $u$ and $v$ lines, then cones become extraordinary vertices of valence different from 4, and holonomy angles on non-contractible loops define how quads are linked together along closed loops.  As this qualitative behavior is a defining quality of a parametrization, its control is critical: e.g., changing one holonomy angle on a loop may increase the distortion considerably, and make alignment with natural directions on the surface impossible.  

In existing methods, holonomy signature is often determined partially or in full from an optimized cross-field, an optimization process for cone placement, or user input (placing cones). To have complete freedom of manipulating parametrization topology it is natural to ask the question on the constraints that need to be observed, specifically

\begin{displayquote}
\emph{For which holonomy signatures, seamless parametrizations with corresponding topology exist?}
\citep{shen2022cross}
\end{displayquote}

Under mild assumptions, this question was answered in 
\citet{shen2022cross}. There is a seamless parametrization of a refinement of an input mesh for \emph{any} holonomy signature (e.g., implied by a cross-field) under a simple condition: it is sufficient to have at least one cone with angle $3\pi/2$ \emph{or} $5\pi/2$ (the actual condition is even weaker), unless the signature has \emph{exactly} two cones ($n_c = 2$) with angles $3\pi/2$ and $5\pi/2$ respectively. 

\citet{shen2022cross} also describes an algorithm which, with an important caveat of numerical precision limitations, produces a seamless parametrization for any valid input holonomy signature. While in principle one can take any input cross-field and produce a parametrization with matching topology, the algorithm is complex and time-consuming, as we discuss in Section~\ref{sec:related} in greater detail. It requires a combinatorial search for suitable loops, and extreme refinement at intermediate stages that may fail at standard floating point precisions.

\paragraph{Contributions.}
We propose a conceptually simple (effectively a Newton iteration on a set of equations), and efficient algorithm that we demonstrate to succeed on a version of Thingi10k dataset, for shapes of widely varying geometric and topological complexity, up to  800,000 vertices and genus over 4,000, as well as on the smaller dataset from \citet{Myles:2014} that includes highly challenging crossfields.  It typically obtains accurate constraint satisfaction and low distortion in just a few iterations. 

A critical element of the algorithm following \citet{capouellez2023metric}, is working in the space of Penner coordinates, which establish a one-to-one correspondence between metrics on meshes with a given set of vertices, and assignment of real numbers to edges, which reduce to logarithmic lengths if an intrinsic Delaunay condition is satisfied. If it is not satisfied, Penner coordinates correspond to valid metrics on different connectivity with the same vertices, obtained by a simple flip algorithm from the original. 

While we do not present a mathematical proof of its correctness, we outline the mathematical reasons why (under suitable limitations on the prescribed angles) we can expect the algorithm to succeed. 

Aside from solving the seamless parametrization problem, the algorithm can be easily modified to solve other types of problems, e.g., the less restricted problem of parametrization with prescribed cones, computing a similarity map with a given holonomy structure, or adding additional angle-based constraints.

\section{Related Work}
\label{sec:related}
There is a wealth of work on various types of parametrization and related geometric problems.  We briefly review the most closely related work; a more complete review can be found in surveys, e.g.,   \citet{naitsat2021inversion} and \citet{fu2021inversion}).  Seamless parametrization in particular is a starting point for many quad layout algorithms, e.g.,  \citet{Bommes:2009,CampenBK15,lyon2021quad} and many others.  A range of general methods that can be applied both for seamless and other parametrizations, assume a feasible starting point already satisfying all constraints, is obtained by another method, e.g., 
\citet{Schueller:LIM:2013,Rabinovich:2017:SLI,liu2018progressive}. 

\paragraph{Intrinsic methods.} Our method belongs to the category of \emph{intrinsic} methods, working with variables intrinsic to the metric, as opposed to representing maps to the plane directly.  Many intrinsic methods were proposed \citep{sheffer2001parameterization, kharevych2006discrete,springborn2008conformal,BenChen:2008}. 
Among these, the most closely related to our work is the approach to discrete conformal maps described in 
\citet{springborn2008conformal} and extended in \citet{campen2021efficient} and \citet{gillespie2021discrete} to methods providing guarantees of correctness,  based on the mathematical foundations built in \citet{gu2018discrete,gu2018discrete2,springborn2019ideal}.
Angle constraints are natural for intrinsic methods; however, so far, no intrinsic method has been proposed for a full set of holonomy constraints: critically, there are not enough degrees of freedom in discrete conformal maps to satisfy all seamlessness constraints.

Another important extension, which considers \emph{similarity maps} extending conformal maps, is \citet{Campen:2017:SimilarityMaps}. While this method provides greater flexibility and thus making it possible to satisfy constraints on the loops, the resulting maps do not correspond to a metric.

Most recently, \citet{capouellez2023metric}, extended these approaches to general metric optimization, which is the closest work to ours.  Crucially, it still relies on conformal maps to enforce constraints, and has the same limitation on the number of constraints that can be enforced. We discuss the similarities and differences in Section~\ref{sec:math} in more detail. 

\paragraph{Seamless parametrization constructions.}
Many methods for seamless parametrization were based on parametric plane coordinates. While most aim to preserve cones and input field topology, i.e., holonomy signature, in many cases, this goal is not stated explicitly. 

To the best of our knowledge,  \citet{shen2022cross} is the only work presenting a method handling general seamless constraints with theoretical guarantees; however, the algorithm has many complex stages, and first constructs an extremely distorted parametrization that needs to be optimized at considerable expense.  Less complete solutions with guarantees include \citet{campen2018seamless}  and \citet{zhou2020combinatorial}  which also involve a highly distorted parametrization stage, and do not provide control over loop holonomy angles. Another work with partial control of holonomy is \citet{levi2023seamless}. 

The majority of methods are used for seamless parametrization, starting with \citet{tong2006dqd,kalberer2007qsp}, often in the context of quadrangulation applications.  These methods often do not guarantee injectivity or finding a solution.  For example,  \citet{Bommes:2009}, which is a foundation for many quadrangulation methods, uses a heuristic change of weights in an optimization problem to reduce the chance of foldovers. Other methods, e.g., \citet{Bommes:2013,CampenBK15,Bright:2017:HGP,hefetz2019subspace} use various types of convexified injectivity constraints
\citep{Lipman:2012}, but there is no guarantee that a solution can be obtained; as shown in \citet{Myles:2014}, these methods do not find a valid solution in many cases. An alternative approach is to construct a T-mesh partition of the surface that does not necessarily correspond to a valid seamless parametrization, e.g., by tracing a cross-field, and then modify it by inserting or merging singularities \citep{Myles:2014,lyon2021quad}. 
In comparison, our method produces a result without failures on a large dataset, with numerical difficulties only for extremely low mesh quality. 

\paragraph{Cross-fields.} As most common holonomy angles for parametrization are determined by a cross-field or a frame field  \citep{Vaxman:FieldsSTAR}, we briefly mention important work on field generation:   \citet{NSymmetry,CraneTrivial,LiPeriodJumps,Bommes:2009,Ray:2009:GeometryAware,farchi2018integer}; many of these offer control over the fields' turning numbers.  Moreover, as demonstrated in \citet{NSymmetry}, a metric field can be obtained for \emph{any} set of turning numbers, the field equivalent of the holonomy signature for parametrizations.  In contrast, for parametrizations there are some exceptions, e.g., signatures with exactly two cones with angles $3\pi/2$ and $5\pi/2$.

\section{Overview}
\label{sec:overview}

We start with a precise formulation of a "naive" algorithm (that, in general, does not work) to make the problem formulation exact, and explain the approach in the simplest form. 

Let $M$ be a triangular mesh with $N_v$ vertices, $N_f$ faces, and $N_e$ edges. For a vertex $i$ and incident triangle $T$, let $\alpha_i^T$ be the angle of $T$ at vertex $i$. 
Consider $2g$ dual loops $L_j$, $j=1\ldots 2g$, i.e., chains of triangles $T^j_m$, with two sequential triangles sharing an edge, and the first and last triangles sharing an edge. Each triangle has exactly one edge on the boundary of the triangle; let $\alpha^j_m$ be the angle opposite this edge. This is the angle between two internal edges $e^j_{m-1}$ and $e^j_m$ of the loop, where $m-1$ is modulo loop length $n_j$. The notation is illustrated in Figure~\ref{fig:notation}.

The \emph{vertex angles} are defined as $\sum_{T \ni i} \alpha^T_i$, where the summation is over all triangles incident at vertex $i$. The loop \emph{holonomy angles} are defined as $\sum_{m=1}^{n_j} d^j_m \alpha^j_m$ where $d^j_m$ is $1$, if the rotation from $e^j_{m-1}$ to $e^j_m$ is counterclockwise, and $-1$ if it is clockwise. This sum is equal to the discrete geodesic curvature of the loop, as each signed angle measures the rotation between two dual edges. 

The \emph{holonomy signature} is an assignment $k^v_i$, $i=1\ldots N_v$, and $k^\ell_j$, $j = 1\ldots 2g$ of 
integers to vertices and loops,  corresponding to angles $k^v_i \pi/2$ and $k^\ell_j \pi/2$, and  satisfying the discrete Gauss-Bonnet theorem, i.e., the sum of all vertex angles should be equal to $2\pi(2-2g)$

For an edge $e$, let $\ell_e$ denote the length of the edge, and $\lambda_e = 2\log \ell_e$ the (scaled) logarithmic edge length. Clearly, the angles can be computed from edge lengths: 
$\alpha^T_i = \alpha^T_i(\lambda)$, where $\lambda \in \bR^{N_e}$ is a vector of the logarithmic edge lengths.

The seamless metric is defined as a (nonunique) solution with respect to $\lambda$ of the following constrained system of $N_v + 2g-1$ equations in $N_e$ variables  (one vertex constraint is redundant due to Gauss-Bonnet):
\begin{figure}[htb]
    \centering    \includegraphics[width=0.95\columnwidth]{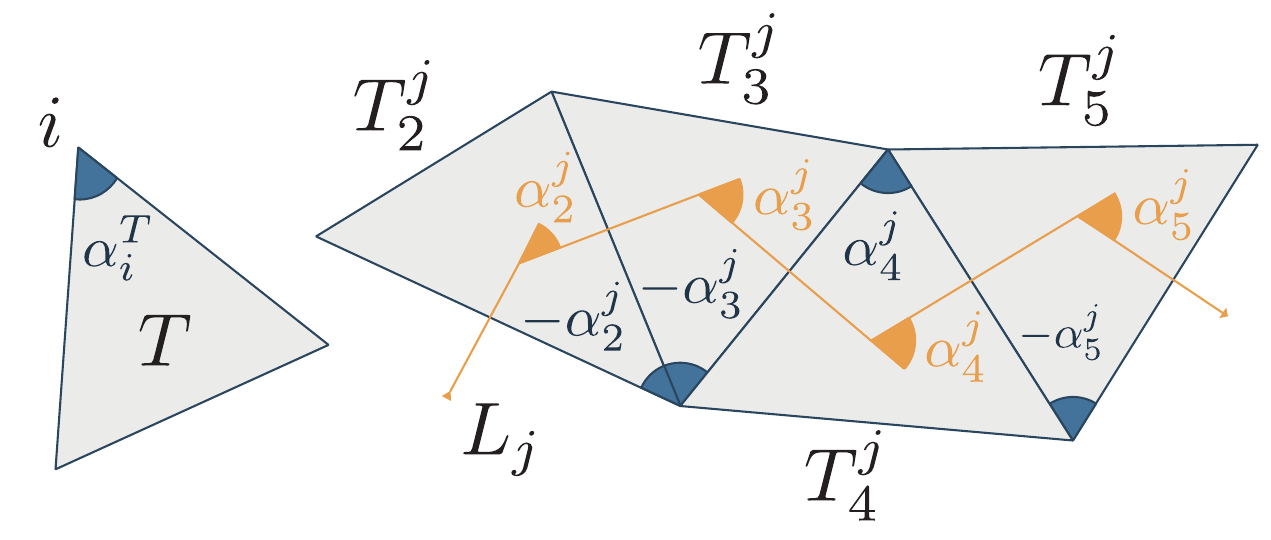}
    \caption{Holonomy angle notation. The signed loop holonomy angles $d_m^j \alpha_m^j$ measure the rotation between dual edges.}
    \label{fig:notation}
\end{figure}
\begin{equation}
\begin{split}
& \sum_{T \ni i} \alpha^T_i(\lambda) = \frac{k^v_i \pi}{2},\,i=1\ldots N_v-1,\;\\
&  \sum_{m=1}^{n_j} d^j_m \alpha^j_m(\lambda) =  \frac{k^\ell_j \pi}{2}, 
\, j=1\ldots 2g.\\
\end{split}
\label{eq:system}
\end{equation}
This is a nonlinear, underdetermined system of equations in variables $\lambda$.  Let $\alpha(\lambda)$ be the vector of triangle angles, and let $C$ be the $(N_v+2g-1) \times (3 N_f)$ matrix of equations above, which are linear in $\alpha$, and  $\Theta$ the vector of holonomy signature angles.  Then in vector form, the system is
$
C\alpha(\lambda) = \Theta
$.

The naive algorithm for obtaining a feasible seamless parametrization is to solve this system of equations using Newton's method (it is not as naive as it sounds, as we explain below).  As the system is underdetermined, we can solve it using the \emph{extended} Newton method.  
Let $F(\lambda) = C \alpha(\lambda) -\Theta$ with $(N_v+2g-1) \times  N_e$ Jacobian matrix $\nabla F = C \nabla_{\lambda} \alpha$, 

\begin{algorithm}[h!]
\setstretch{0.9}
\SetAlgoLined
\DontPrintSemicolon
\SetKwInOut{Input}{Input}
\SetKwInOut{Output}{Output}
\SetKwProg{Fn}{Function}{:}{}
\SetKwRepeat{Do}{do}{while}
\SetKw{Not}{not}
\Fn{\scp{FeasibleSeamless}$(\lambda^0,\Theta)$}{
$\lambda \gets \lambda^0$\;
\While{ \Not{\scp{Converged}$(\lambda,\Theta)$}}{
     $\DF \gets C \nabla_{\lambda}\alpha $\;
     $L \gets  \DF  \DF^T$\;
     Solve $L \mu   = -F$\;  
     $d \gets \DF^T \mu$\;
     $\beta \gets$ \scp{LineSearch}$(\lambda, d)$\; 
     $\lambda \gets \lambda + \beta d$.
    }
    \Return $\lambda$ 
}
\caption{Naive Newton algorithm.}
\label{alg:naive}
\end{algorithm}
We note that the gradient $\nabla_\lambda \alpha$ is closely related to the well-known cotangent matrix.

It is important that the algorithm is initialized with $\lambda^0$,  the original edge lengths.  If the constraints were linear, then the pseudoinverse solve would minimize the norm  $\|\lambda - \lambda^0\|^2$, which is a measure of isometry used in \citet{capouellez2023metric}.  In the nonlinear setting, the change in the norm is minimized at each step; this is not equivalent to minimizing the norm subject to linear constraints, but, as we will see, it is a useful approximation. 

However, clearly, there is no \emph{a priori} reason to believe that the algorithm can find a solution to the constraint system: the solution may not even exist on fixed connectivity.  In the next sections, we make changes to the algorithm which ensures that it converges in the case of seamless parametrization with only vertex constraints and provide empirical evidence and mathematical considerations that it can also handle the additional loop constraints. 

\paragraph{Connection to discrete conformal maps.} We further note that the algorithm is not as naive as it may first appear.  With two changes, (1) eliminating the constraints on loops, and (2) reducing variables $\lambda$  to $N_v-1$ vertex logarithmic scale factors $u$ via $\lambda_e = \lambda_e^0 + u_i + u_j$, where vertices $i$ and $j$ are the endpoints of edge $e$, the $N_v-1$ \emph{vertex} constraints
happen to be the gradient of a convex function with respect to scale factors $u$. As a consequence, solving this restricted version of the constraint system by the Newton method is efficient and robust, and globally converges to a solution \emph{if one exists}, as demonstrated in \citet{campen2021efficient,gillespie2021discrete}.  
The critical step for making the conformal algorithm provably convergent was to enlarge the space by allowing connectivity changes, which we do next for our algorithm.

\section{Background: Penner coordinates}
\label{sec:background}

We briefly summarize the idea and use of Penner coordinates.  While the type of problems we consider (seamless parametrization with prescribed cones) does not necessarily have a solution for an arbitrary input mesh connectivity, it turns out that it can be solved if the connectivity is allowed to change, and it is sufficient to change it in a restrictive way: specifically, the vertices remain the same, but the connectivity may change through edge flips. However, performing optimization on varying mesh connectivity is difficult, as the variables and equations typically are connectivity-dependent.  \emph{Penner coordinates} provide a way to parameterize, with coherent variables, \emph{all} metrics defined on \emph{all} mesh connectivities sharing the same vertex set. 

We start with an assignment $(M, \ell)$ of edge lengths $\ell$ to the edges of a mesh $M$, satisfying the triangle inequality.  We will also use $(M,\lambda)$ to denote an equivalent assignment of logarithmic lengths.
(We use logarithmic edge lengths to eliminate a positivity constraint, and as we will see these are particularly natural for our algorithm.)

\begin{figure}
    \centering
    \includegraphics{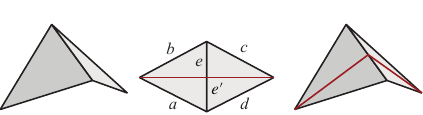}
    \caption{Intrinsic flip. Week's flip algorithm uses Ptolemy formula for $\ell(e')$ which coincides with the Euclidean length if the angles opposite to $e'$ sum up to $\pi$.}
    \label{fig:flip-notation}
\end{figure}

If we consider each triangle as flat, this assignment defines a \emph{cone metric} on the mesh, with nonzero curvature only at vertices. If we are allowed to change connectivity, then there are many ways to describe the same metric: if we perform an intrinsic edge flip (Figure~\ref{fig:flip-notation}), the metric does not change, but we get new connectivity and new edge lengths $(M', \ell')$. 

We can convert a description of a metric in terms of edge lengths to a (nearly)  uniquely defined one if we require that the mesh $M$ is \emph{Delaunay}. In other words, given a pair $(M,\ell)$, we can produce a pair $(\tM,\tilde{\ell})
=\del(M,\ell)$ such that each edge $e$ of $\tM$ satisfies the intrinsic Delaunay condition ${\alpha}_i + {\alpha}_j \leq \pi$, where ${\alpha}_i,{\alpha}_j$ are the triangle angles opposite edge $e$. In terms of edge lengths, the Delaunay condition for edge $e$ is equivalent to a simple rational expression:
\begin{equation}
\frac{\ell(a)^2 + \ell(b)^2-\ell(e)^2}{2\ell(a)\ell(b)} + 
\frac{\ell(c)^2 + \ell(d)^2-\ell(e)^2}{2\ell(c)\ell(d)} \geq 0 
\label{eq:ideal-delaunay}
\end{equation}
$\del(M,\ell)$ can be computed using the standard intrinsic Delaunay flip algorithm, which repeatedly flips every edge that does not satisfy this condition until none are left.

If we fix the connectivity $M$, then there is a set of choices of cone metrics for which $M$ is Delaunay.  This set of metrics is called a \emph{Penner cell} of $M$, which we denote $\cP(M)$.
For each cell, we have local coordinates for metrics, which are simply the edge lengths $\ell$.

\begin{figure}
    \centering
    \includegraphics[width=0.6\columnwidth]{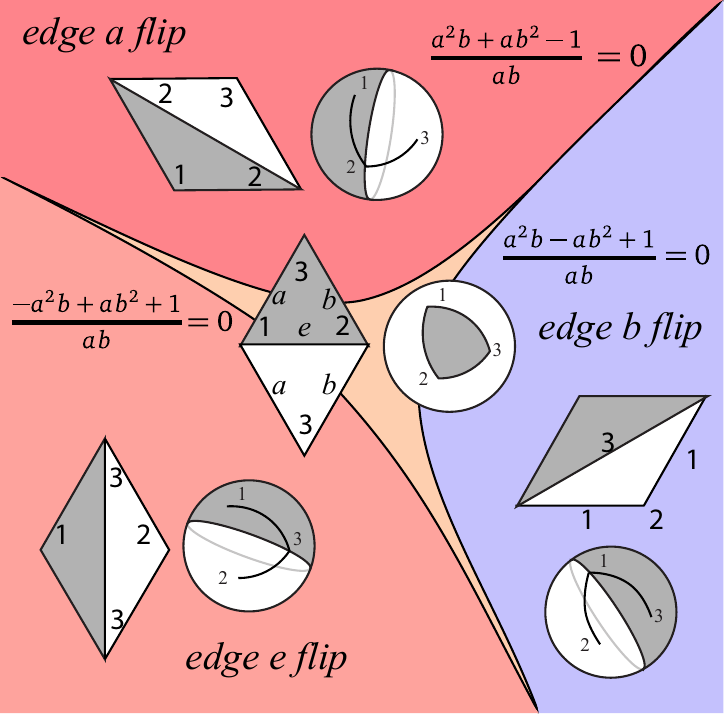}
    \caption{Penner cell decomposition of cone metrics with 3 vertices and 3 edges (Figure from \citet{capouellez2023metric})}
    \label{fig:penner-cells}
\end{figure}

Penner cells cover the whole space of metrics with a fixed set of vertices and surface genus (Figure~\ref{fig:penner-cells}). Two adjacent cells differ by a single flip at an edge $e$. Since the sum of angles opposite $e$ is equal to $\pi$ at the boundary between the cells, this flip corresponds to a simple transformation for the lengths, defined by the Ptolemy formula.  Removing $e$ and inserting  the flipped edge $e'$ in a pair of adjacent triangles with external edges  $\ell_a,\ell_b,\ell_c,\ell_d$  corresponds to the edge length update $
\ell'(e') = \frac{\ell(a)\ell(c) + \ell(b)\ell(d)}{\ell(e)}, 
$
and  $\ell'(f)  = \ell(f)$ for all edges $f \neq e$. These formulas define a transition from lengths with respect to connectivities $M$ and $M'$ of two adjacent cells. We denote this transition map $\tau(\ell)$.

The idea of Penner coordinates is to extend the length coordinates $\ell$ on one, arbitrarily picked, cell $\cP(M_0)$ to the whole space of metrics, by using the formulas above as transition maps. Note that the formula can be applied to an assignment of positive numbers to the edges whether these correspond to actual lengths (i.e., satisfy triangle inequality) or not. Then for a chain of flips of edges $e_1,...,e_n$ connecting two connectivities $M$ and $M_0$, we can define the transition map $\tau(M,M_0): \bR^{N_e}_+ \to \bR^{N_e}_+$ as the composition $\tau_{n} \circ \tau_{n-1} \circ \cdots \circ \tau_1$ of the transition maps for the individual flips. These transition maps are smooth and well-defined, i.e., they do not depend on the sequence of flips used to construct the map \citep{penner1987decorated}.

More formally, given a connectivity $M_0$ and a metric $(M,\ell)$ the Penner coordinates of $(M,\ell)$ are defined as follows: 

\begin{definition}
\citep{capouellez2023metric}
Penner coordinates for a cone metric with length coordinates $(M,\ell)$ in Penner cell $\cP(M)$, with respect to
$M_0$ is a vector $P_{M_0}(M, \ell)$ of positive numbers in $\bR^{N_e}_+$ defined as  
$
P_{M_0}(M, \ell)  = \tau(M,M_0)(\ell).$
i.e., simply the coordinate change from $\cP(M)$ to $\cP(M_0)$ by a composition of Ptolemy formulas.  
\end{definition}

The key features of Penner coordinates that we need for our algorithm are:

\begin{itemize}
    \item For a fixed mesh $M_0$, any choice of logarithmic edge length assignments defines a metric. Its canonical representation $(\tM,\tl)$ can be obtained by the Week's flip algorithm, which is identical to the standard Delaunay flip algorithm, but with length updates based on the Ptolemy formula.  It is important to note that the Delaunay criterion expressed in terms of lengths is well-defined for Penner coordinates, i.e., does not require triangle inequality, and Week's algorithm is guaranteed to produce a Delaunay mesh, i.e., at termination, Penner coordinates become lengths.
    \item  Conversely, for any metric $(M,\lambda)$ its logarithmic Penner coordinates $(M_0,\lambda^0)$ can be obtained by finding a flip sequence connecting $M$ and $M_0$ and applying Ptolemy transformations to $\lambda$ at ever flip. 
\end{itemize}
In this way, (logarithmic) Penner coordinates establish a one-to-one correspondence between the space of metrics on a mesh and $\bR^{N_e}$.

The idea of extending Algorithm~\ref{alg:naive} to work on arbitrary meshes is to apply it to Penner coordinates on a mesh, which allows one to optimize in a larger space where,  e.g., solutions for conformal maps are known to exist.

\section{Complete Algorithm}\label{sec:method}

Next, we describe the modified version of Algorithm~\ref{alg:naive}, now extended to solving the system over the whole space of metrics.  
Compared to the naive version, the most significant change to the algorithm is that the optimization is performed in Penner coordinates, i.e., variables assigned to the edges of the initial connectivity, which may not satisfy the triangle inequality.

To compute the angles that are needed both for constraints and constraint Jacobians in the algorithm, we simply apply the flip algorithm with Ptolemy length changes, to find the connectivity $\tM$ which is Delaunay with respect to the updated coordinates $\tl$. 
As $\tM$ is Delaunay w.r.t, $\tl$, the triangle inequality is satisfied, and angles can be computed. Our constraint function $F$ in Penner coordinates thus becomes
\begin{equation}
F(\lambda) =  C \alpha(\del(M_0,\lambda)) - \Theta = 0
\label{eq:penner-constraints}
\end{equation}

The rest of the algorithm is largely unchanged, but some additional work is also necessary to update the angle Jacobian $\nabla_\lambda \alpha$ and the constraint matrix $C$ in Penner coordinates, which we now delineate.

\paragraph{Jacobian update.} In order to compute 
\[
\nabla_\lambda \alpha = \nabla_{\tl} \alpha \cdot \nabla_\lambda \del,
\]
we need to compute the Jacobian $\nabla_\lambda \del$ of the transition map to the Delaunay connectivity $\tM$. We use an incremental update of the Jacobian matrix as in \citet{capouellez2023metric}.  Specifically, the function \scp{DiffPtolemy}$(M,\ell, e)$  computes an update matrix corresponding to the Jacobian of the transition map corresponding to the change of coordinates resulting from flipping the edge $e$, with incident edges $a,b,c,d$ as in Figure~\ref{fig:flip-notation}. Define the standard shear
$$
t = \frac{\ell_a \ell_c}{\ell_b\ell_d}.
$$
The matrix of derivatives of the transition map with respect to $\lambda$, 
is an identity matrix, except the rows corresponding to $e$, 
which is zero except the subrow corresponding to  edges $e,a,b,c,d$, which has the form:
\[
D_{e,[e,a,b,c,d]} = 
\left[-2,
\frac{2t}{1+t}, \frac{2}{1+t}, 
\frac{2t}{1+t}, \frac{2}{1+t}, 
\right]
\]

By the chain rule, $\nabla_\lambda \del$ is simply the product of the matrices corresponding to the sequence of flips determined by the flip algorithm.

\begin{algorithm}[b]
\setstretch{0.9}
\SetAlgoLined
\DontPrintSemicolon
\SetKwInOut{Input}{Input}
\SetKwInOut{Output}{Output}
\SetKwProg{Fn}{Function}{:}{}
\SetKwRepeat{Do}{do}{while}
\SetKw{Not}{not}
\Input{
    triangle mesh $M = (V,E,F)$, closed, manifold,\newline
    edge lengths $\ell^0 = e^{\lambda^0/2}$ satisfying triangle inequality,\newline
    target angles $\hat\Theta > 0$ respecting Gauss-Bonnet at vertices, and on a basis of dual loops.
}
    \vspace{2pt}
\Output{
    triangle mesh $\tM = (V,\tilde{E},\tilde{F})$,\newline
    edge lengths $e^{\tl / 2}$ satisfying triangle inequality,\newline 
    with angles $\max_\Theta \|\Theta - \hat\Theta\| \leq \epsilon_c$.
    }
    \vspace{2pt}
\Fn{\sc{FeasibleSeamless}$(M,\lambda^0,\hTh)$}{
$\lambda \gets \lambda^0$ \;
\While{ \Not \scp{Converged}$(M,\lambda)$}{
    $\tM, \tl ,D, C  \gets$  \scp{DiffMakeDelaunay}$(M,\lambda,C)$\;
    $\alpha,  \nabla_{\tl}\alpha \gets$ \scp{ComputeAnglesAndGradient}$(\tM,\tl)$ \;
     $\DF \gets C\, \nabla_{\tl}\alpha\, D$\;
     $L \gets  \DF  \DF^T$\;
     Solve $L \mu   =  -F$\;  
     $d \gets \DF^T \mu$\;
     $\beta \gets$ \scp{LineSearch}$(\lambda, d)$\;
     $\lambda \gets \lambda + \beta d$
    }
    \Return \scp{MakeDelaunay}$(M,\lambda)$ 
}
\vspace{1ex}
\Fn{\scp{DiffMakeDelaunay}$(M, \lambda,C)$}{
    $\tM,\tl \gets M,\lambda$ \;
	$D \gets \mathrm{Id}$ \;
	$Q  \gets \{e | \mbox{\scp{NonDelaunay}}(M,\lambda, e)\}$ \; 
	\While{$Q  \neq \emptyset$}{
		remove $e$ from $Q$\;
		$\tM,\tl \gets$ \scp{PtolemyFlip}$(\tM,\tl,e)$ \;
		$D \gets$ \scp{DiffPtolemy}$(\tM,\tl, e) \cdot D$ \;
            $C \gets$ \scp{UpdateConstraints}$(\tM,\tl,C,e)$
  }
\Return $\tM,\tl,D,C$
}
\caption{Seamless parametrization algorithm.}
\end{algorithm}

\begin{figure}[t!]
    \centering
    \includegraphics[width=0.8\columnwidth]{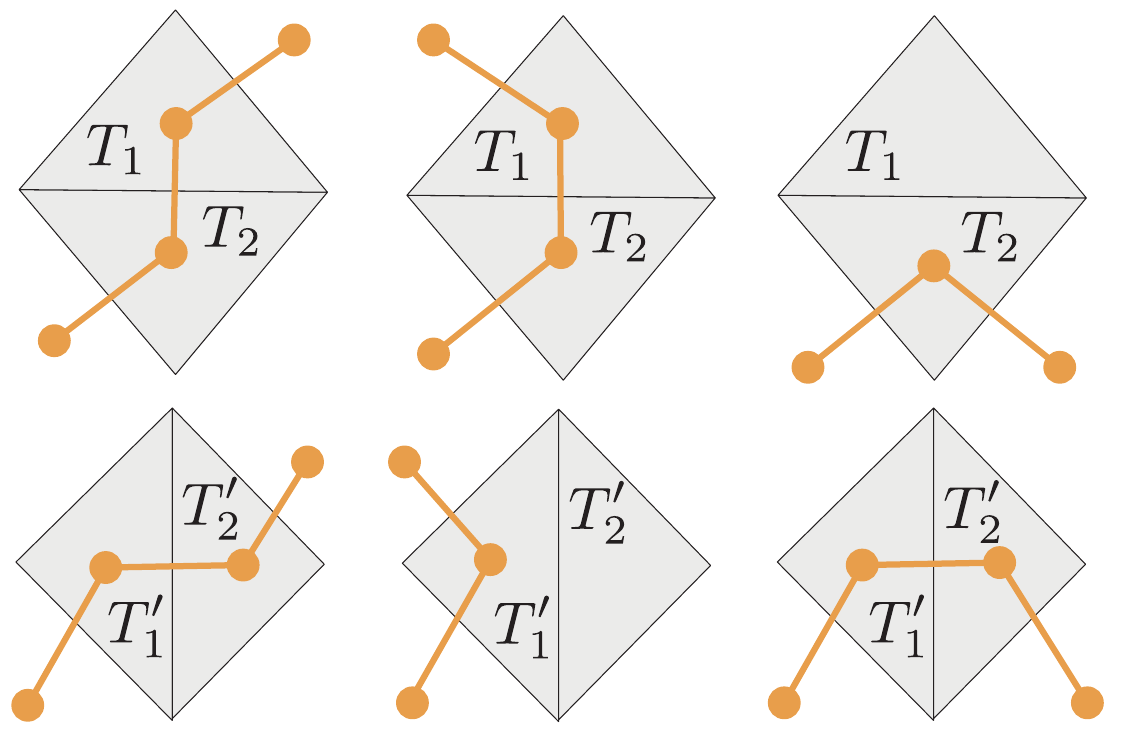}
    \caption{Dual loop update for a flip.}
    \label{fig:loop-flip}
\end{figure}

\paragraph{Constraint update.}  Another complication that needs to be addressed is the update of the constraints themselves, which must be formulated in the Delaunay connectivity $\tM$ where we can compute angles. 
The update of the vertex angle constraints is simple: the angles incident at the vertex in $\tM$ are used for the constraint; the update for the loops is more complicated.

During each intrinsic flip and for each dual loop $L_j$, we find a dual loop on the flipped mesh $M'$ that is topologically equivalent to the dual loop on the original mesh $M$. More formally, we find a loop $L_j'$ that is homotopic to the original loop $L_j$, in the sense that the closed loop of dual edges defined by $L_j$ can be continuously deformed to that of $L_j'$, without passing through cones. Since the intrinsic metric is flat away from cones, the holonomy of these two dual loops with respect to the metric (which is unchanged by the intrinsic flip) will be the same.

If $L_j$ does not intersect the pair of adjacent triangles $T_1, T_2$ where a flip occurs, then there is nothing to do and the loop is left unchanged. If a dual loop does contain $T_1$ or $T_2$, we locally modify it so that it enters and exits the flipped triangle pair $T_1', T_2'$ across the same edges as in the original mesh. Three representative cases are illustrated in Figure~\ref{fig:loop-flip}. The constraint matrix is then recomputed for the new triangulation; the function
{\sc UpdateConstraints} returns the updated matrix.

\paragraph{Line search.} We use a backtracking line search. Since our extended Newton method does not correspond to a (known) convex energy, we cannot use a sufficient decrease condition, so instead we use the following conditions for the step size $\beta$:
\begin{enumerate}
    \item The norm of the constraint vector does not increase,
    \[\|F(\lambda + \beta d)\| \leq \|F(\lambda)\|\]
    \item The direction of the constraint vector does not reverse,
    \[F(\lambda) \cdot F(\lambda + \beta d) \geq 0 \]
\end{enumerate}
The first condition ensures that we make global progress in each iteration toward satisfying our constraints, and the latter condition ensures that our descent direction remains a descent direction at the end of the line step.

\paragraph{Postprocessing: overlay meshes.} Once the final lengths on the mesh are computed, 
these need to be mapped to a refinement of the original connectivity (the overlay mesh). We use the approach of \citet{capouellez2023metric} to do this without changes.  Briefly, the intersection points of the original edges and flipped edges are tracked through the flips. As the Penner coordinates in intermediate configurations do not necessarily satisfy the triangle inequality, hyperbolic geometry considerations are used to assign edge coordinates to these vertices.  At the end of the process, lengths are assigned to the edges of the overlay mesh, and it is mapped to the plane using a layout process similar to \citet{springborn2008conformal}. 
The process is fast and robust. Whenever possible, inserted vertices are removed, resulting only in a small percentage increase in mesh size relative to the original \citep{capouellez2023metric}.

\paragraph{Preprocessing: intrinsic improvement.} Since our method is fundamentally intrinsic, it is amenable to intrinsic preprocessing to improve the initial triangle quality without modifying the original geometry. Intrinsic Delaunay refinement \cite{sharp2019navigating}, which inserts vertices in triangles with small angles to improve triangle quality while maintaining the Delaunay property with edge flips, is a particularly natural choice. Such vertex insertion is guaranteed to produce meshes with a minimum triangle angle of up to $30^{\circ}$ at almost all vertices; however, thin needle-like features (see inset) with a very small total angle at a vertex cannot be improved by such intrinsic refinement. We use the implementation provided by \citet{geometrycentral} with vertex insertions sufficient to produce triangle angles all above some threshold $\alpha_{\min}$ (away from needle-like features).

\begin{wrapfigure}{l}{0.1\columnwidth}
    \centering
    \includegraphics[width=0.1\columnwidth]{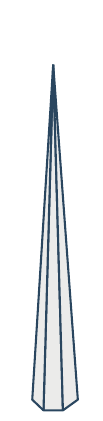}
    \label{fig:needle}
\end{wrapfigure}

A simpler complimentary approach for intrinsic preprocessing is to interpolate between the Penner coordinates $\lambda^0$ of the original metric and perfectly regular Penner coordinates $\lambda = 0$, which corresponds to a metric of completely equilateral triangles with a uniform angle of $60^{\circ}$. In other words, rather than initializing our method with $\lambda \gets \lambda^0$, we can use $\lambda \gets \beta \lambda^0$ for any $\beta \in [0, 1]$. Such interpolation can achieve arbitrarily regular triangles, but there is a trade-off between the initial triangle quality and the final geometric distortion as $\lambda \approx 0$ may initialize the optimization with a metric already far from $\lambda^0$.

We use a line search $\lambda^0 = \beta^n \lambda$ with $\beta = 0.9$ until $\min\alpha(\lambda^0) \geq \alpha_{\min}$. We also recenter $\lambda^0$, which corresponds to a global scaling with no impact on $\alpha$, after interpolation so that the average Penner coordinate does not change, i.e., $\frac{1}{N_e}\sum_{e \in E} \lambda_e^0 = \frac{1}{N_e} \sum_{e \in E} \lambda_e$. This allows us to partially reduce the geometric distortion of the interpolation without compromising triangle quality.

\section{Mathematical considerations}\label{sec:math}

The empirical performance of the algorithm shown in Section~\ref{sec:eval} provides strong evidence that the algorithm is likely to be related to a convex optimization problem: it is highly unlikely that Newton method in a vast majority of cases converges in a near-optimal number of iterations, and in all tested cases converges to a solution (with some caveats for extremely poor quality meshes).  At the same time, three closely related problems do correspond to convex problems: (1) metric optimization of convex objectives described in \citet{capouellez2023metric} subject to angle constraints at vertices, but not on dual loops, (2) the conformal mapping based on scale factors described in \citet{campen2021efficient,gillespie2021discrete}, and (3) the similarity mapping based on a scale-factor 1-form $\psi$ described in \citet{Campen:2017:SimilarityMaps}, with details on the convex functional provided in \citet{Campen:2017:OnSimilarityMaps}.

\paragraph{Comparison to metric optimization.} Our algorithm is very close to the projected gradient algorithm of \citet{capouellez2023metric}, but is different in critical respects. 
The most important difference is the \emph{absence of conformal projection}.  The algorithm of \citet{capouellez2023metric} performs a conformal projection at every state to enforce constraints, which requires an inner loop conformal solve, and more importantly limits the supported constraints to the ones conformal maps support (i.e., not full seamless constraints).  

The second major difference is the absence of explicitly optimized energy. 
While this limits the type of parametrizations the method can produce, at the same time, this has a substantial impact on the algorithm's performance.  Note that the method described in \citet{capouellez2023metric} is inherently a first-order method, with linear convergence.  To apply a second-order Newton method in the setting of constrained optimization, would require second derivatives of the constraints.  However, these are known to be discontinuous.  At the same time, our method is a Newton method with quadratic convergence, and only \emph{first} derivatives of constraints are needed. 

\paragraph{Comparison to conformal mapping.} As described in Section~\ref{sec:overview}, conformal methods, e.g., \citet{campen2021efficient,gillespie2021discrete}, operate in a reduced subspace of Penner coordinates spanned by logarithmic scale factors $u \in \bR^{N_v}$. They minimize the convex function
\[
E(u) = \sum_{T \in \tM} \left(2f(\tl_{a},\tl_{b},\tl_{c}) - \pi(u_i+u_j+u_k)\right) + \Theta_V^\intercal u
\]
where $a, b, c$ and $i, j, k$ are the edges and vertices of triangle $T'$ respectively (see inset), and $\Theta_V$ are the vertex angle constraints.

\begin{wrapfigure}{l}{0.25\columnwidth}
    \centering
    \includegraphics[width=0.275\columnwidth]{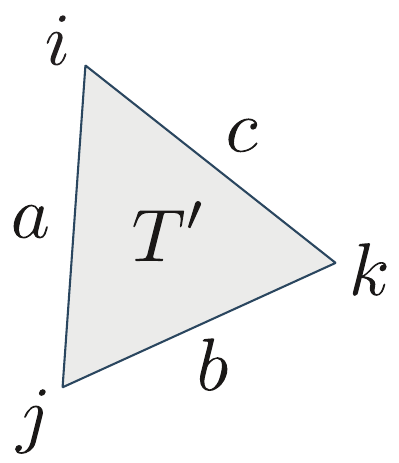}
    \label{fig:conformal}
\end{wrapfigure}

As in our method, $(\tM, \tl) = \del(M, \lambda)$, where $\lambda_{a} = \lambda_a^0 + u_i + u_j$. $f$ is a per-triangle function involving Milnor's Lobachevsky function \citep[Eq.~(8)]{springborn2008conformal} with the important property that
\[
\frac{\partial  f}{\partial \tl_{a}} = \alpha_k^{T},
\]
and the gradient of $E(u)$ is precisely the vertex constraints.

The iteration of our algorithm is very similar to the Newton iteration used to solve for conformal maps in \citet{gillespie2021discrete} and \citet{campen2021efficient}. There are several important differences: most significantly, as we use $N_e$ length variables, rather than $N_v$ vertex scale factors, we can introduce $V + 2g-1$ holonomy constraints, with degrees of freedom to spare.  As the optimization problem in the conformal case is fully constrained, there is no additional objective optimized directly or indirectly. In our case, each iteration optimizes the deviation from the metric at the previous step.  For this reason, as in \citet{capouellez2023metric}, we evaluate the descent direction and the Jacobian of constraints in Penner coordinates, i.e., with respect to the initial mesh connectivity.

\paragraph{Comparison to similarity mapping.} Similarity methods, e.g., \cite{Campen:2017:SimilarityMaps}, use the energy
\[
E(\psi) = \sum_{T \in M}g(\psi_i^T, \psi_j^T, \psi_k^T) - \Theta^\intercal P^{+} \psi
\]
where $\psi_i^T$ is the value of the closed one-form $\psi$ on the edge of triangle $T$ opposite vertex $i$, and $P^+$ is the pseodoinverse of a matrix formed by coefficient vectors for a basis of closed one-forms. Like $f$, the function $g$ is a triangle function that depends on Milnor's Lobachevsky function with the corresponding property that
\[
\frac{\partial g}{\partial \psi_i^T} = \alpha_i^T.
\]
Similarity map problems were also considered in a similar context in \citet{rivin1994euclidean}. While these methods were developed in a fixed connectivity setting and a fully rigorous analysis using Penner coordinates is still absent in the literature, the theoretical extension is straightforward.

The mathematical approach used successfully to obtain variational principles for seamless constraints do not directly apply in the metric setting since the similarity methods produce a discrete scaling one-form that generally will not be exact and thus not integrable on the surface. However, we note that, intuitively, there are more than enough of degrees of freedom ($N_e$) in the metric $\lambda$ to satisfy $(N_v-1) + 2g$ constraints, and the $N_v - 1$ vertex constraints can be satisfied using the conformal degrees of freedom alone.

\paragraph{Invalid signatures} Unlike in the similarity setting, in the metric setting there are holonomy signatures that theoretically cannot be satisfied by any seamless parameterization. The known invalid signatures are:

\begin{itemize}
    \item any signature with exactly two cones of angles $3\pi/2$ and $5\pi/2$, and
    \item a signature with no cones but nontrivial dual loop holonomy angles, i.e., $k_j^{\ell} \neq 0$ on simple loops. 
\end{itemize}
The first case is considered in \cite{jucovivc1973theorem,izmestiev2013there} and the second follows from the classification of flat metrics on a torus which all correspond to periodic tilings of the plane by parallelograms, and have trivial holonomy.  

By the discrete Gauss-Bonnet theorem, these particular invalid cone prescriptions are only possible on genus 1 surfaces. They can also be interpreted in our metric setting. For instance, signatures with no cones correspond to flat tori, and the holonomy of any dual loop on a flat torus is necessarily trivial, so the loop holonomy constraints are redundant with the vertex angle constraints.

Verifying that our algorithm produces a valid solution whenever one exists, or identifying cases when it does not work is an important future direction.

\section{Evaluation}\label{sec:eval}

\begin{figure}
    \centering
    \includegraphics[width=0.9\columnwidth]{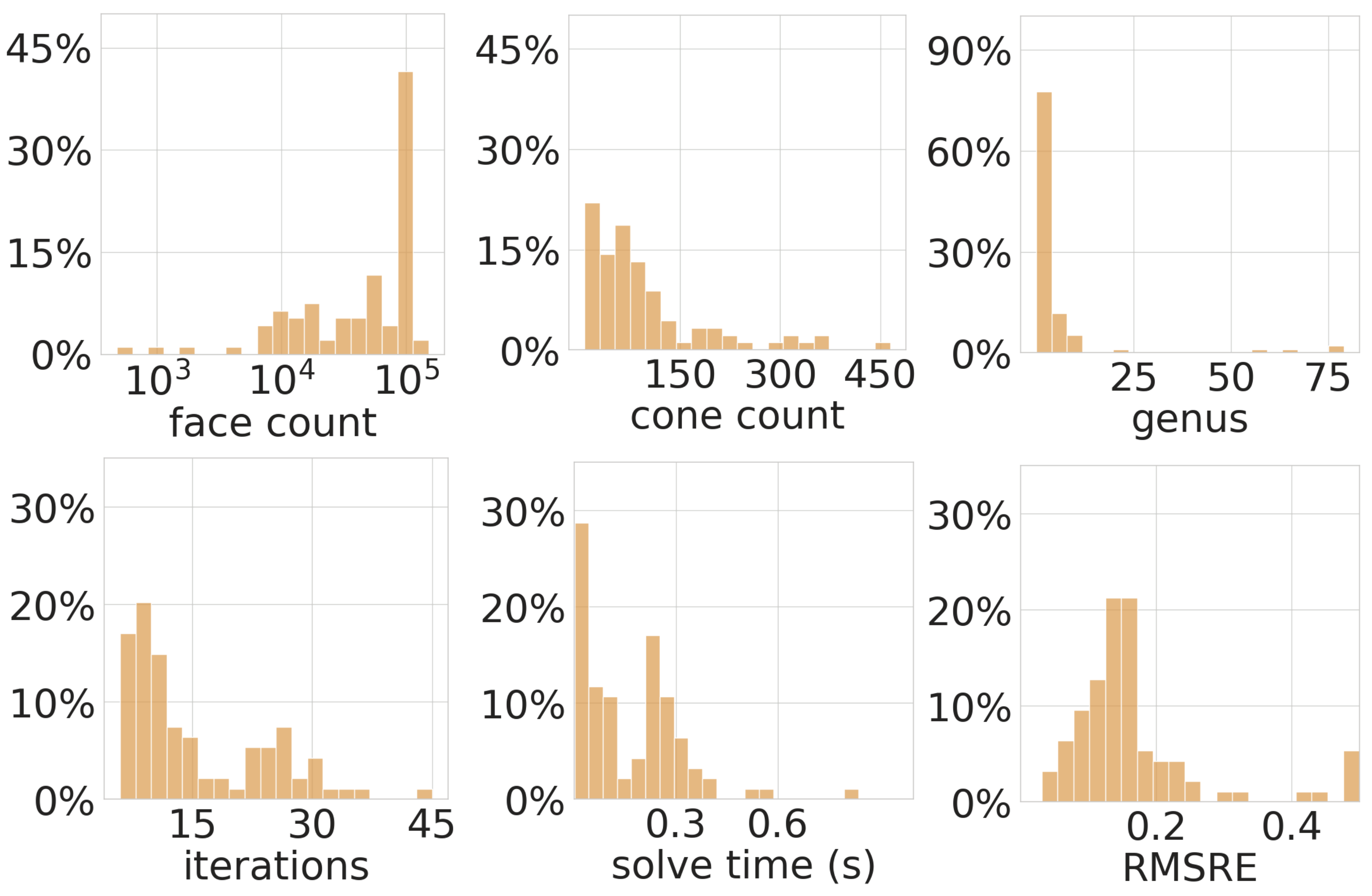}
    \caption{Top: Distributions of mesh face counts, cross-field cone counts, and surface genus for the dataset of \citet{Myles:2014}. Bottom: Distributions of iteration counts, average linear solve times, and RMSRE errors for our method. Outliers are aggregated in the rightmost bin.} 
    \label{fig:myles-dataset}
\end{figure}

\begin{figure*}[t!]
    \centering
    \includegraphics[width=0.9\textwidth]{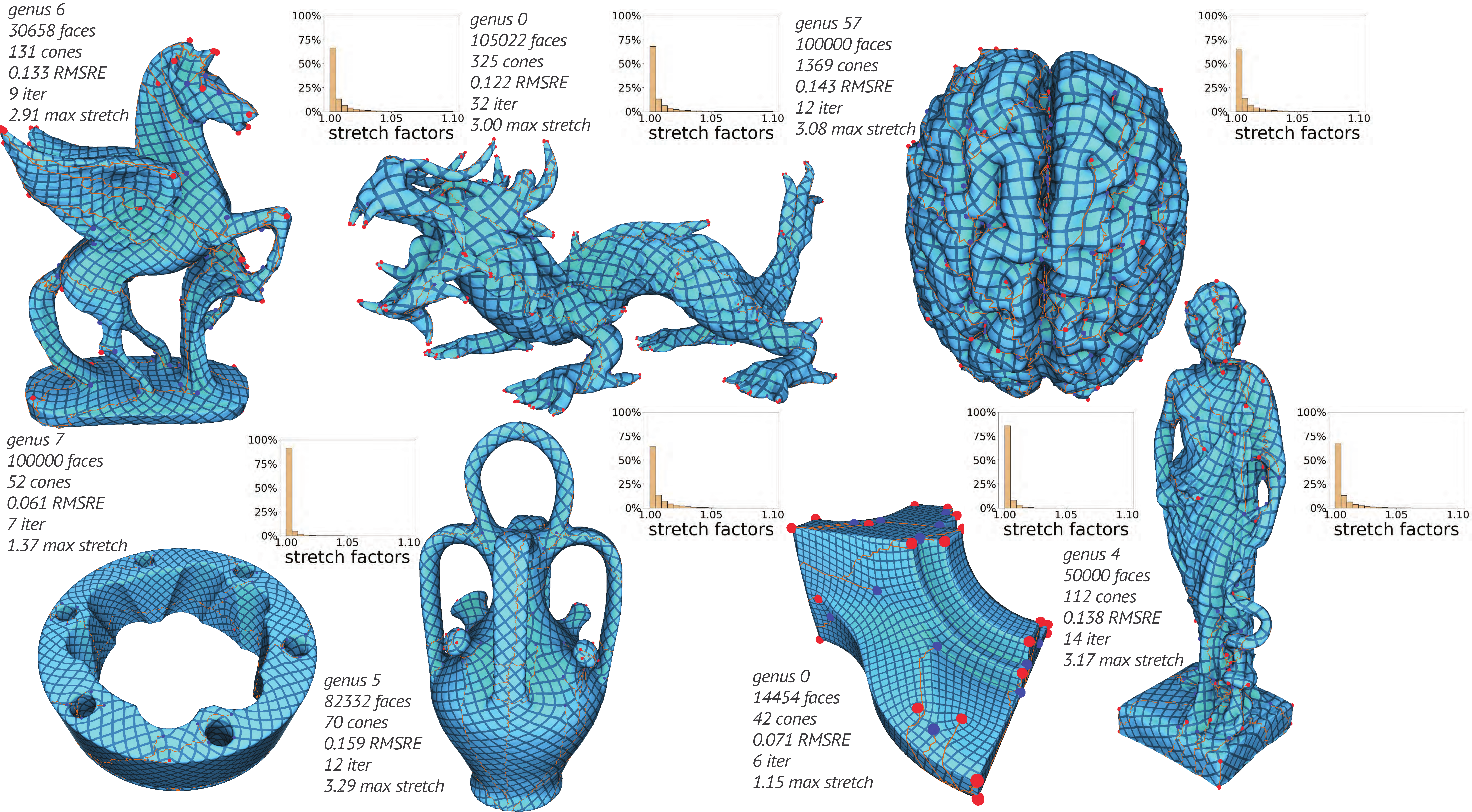}
    \caption{Parametrizations produced by our method for meshes from the dataset of \cite{Myles:2014} with histograms of per-edge symmetric stretch factors.}
    \label{fig:myles-examples}
\end{figure*}

We evaluated our method on 94 closed meshes with challenging fields provided in the dataset of 
\citet{Myles:2014}. Our method produces a parameterization satisfying both vertex and loop holonomy constraints without exception in under fifty iterations, with the main bottleneck being a sparse linear solve for the descent direction, and the resulting parametrizations exhibit low geometric distortion. Example parametrizations are provided in Figure~\ref{fig:myles-examples}. Intrinsic preprocessing is only necessary for three meshes with poor triangle quality, and the final geometric distortion is qualitatively low. We also measure the distortion quantitatively in terms of Penner coordinates using the Root Mean Squared Relative Error
\[
RMSRE(\ell, \ell^0) = \left(\sum_{e \in E}\frac{1}{|E|} \left(\frac{\ell_e - \ell^0_e}{\ell^0_e}\right)^2\right)^{1/2},
\]
where $\ell^0$ are the lengths of the original embedding metric. Distributions of the dataset constraints and the performance of our method are provided in Figure~\ref{fig:myles-dataset}.

\begin{figure}
    \centering
    \includegraphics[width=0.9\columnwidth]{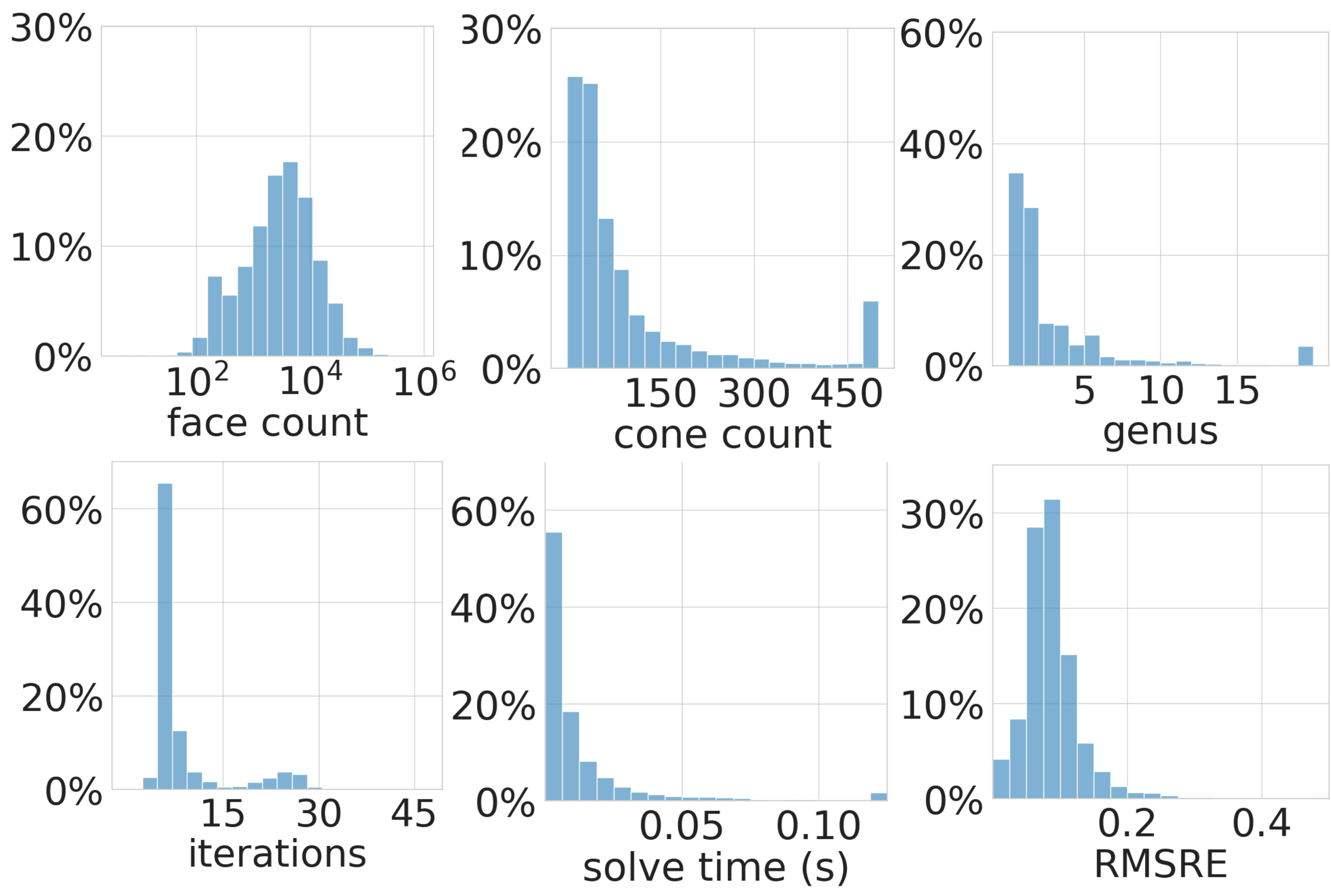}
    \caption{Modified Thingi10k dataset and result statistics. 548 meshes have genus above 20, with a maximum value of 4307.}
    \label{fig:tetwild-dataset}
\end{figure}

\begin{figure*}[t!]
     \centering
     \includegraphics[height=\textheight]{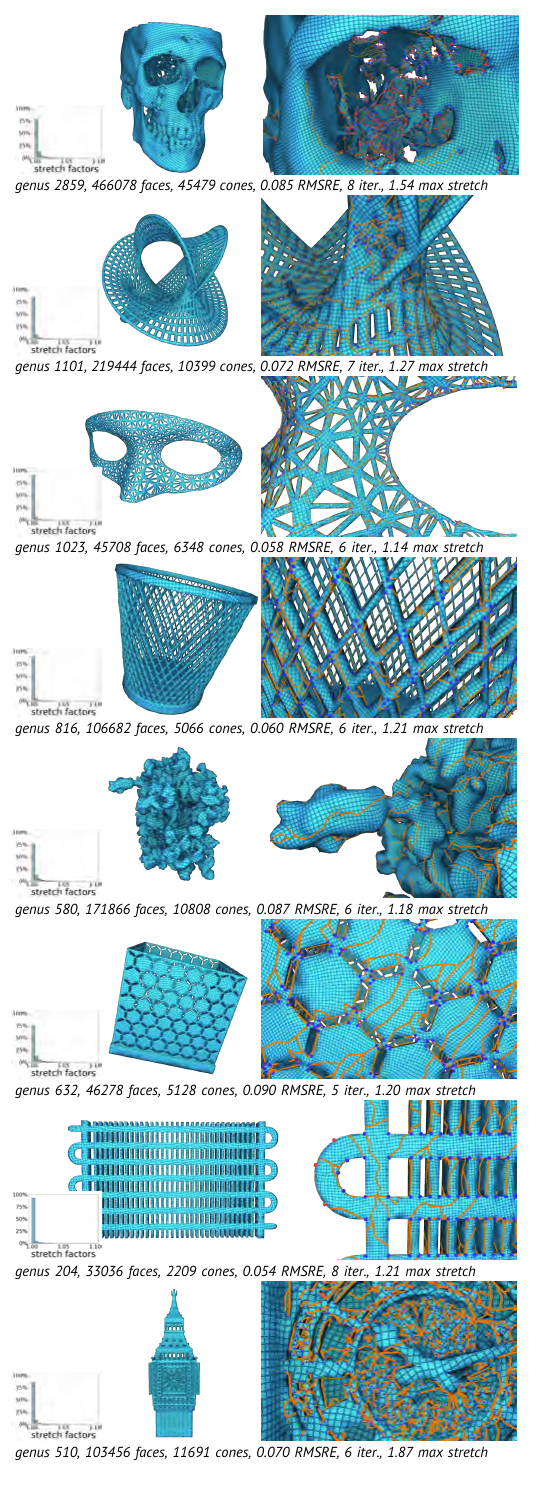}
    ~\includegraphics[height=\textheight]{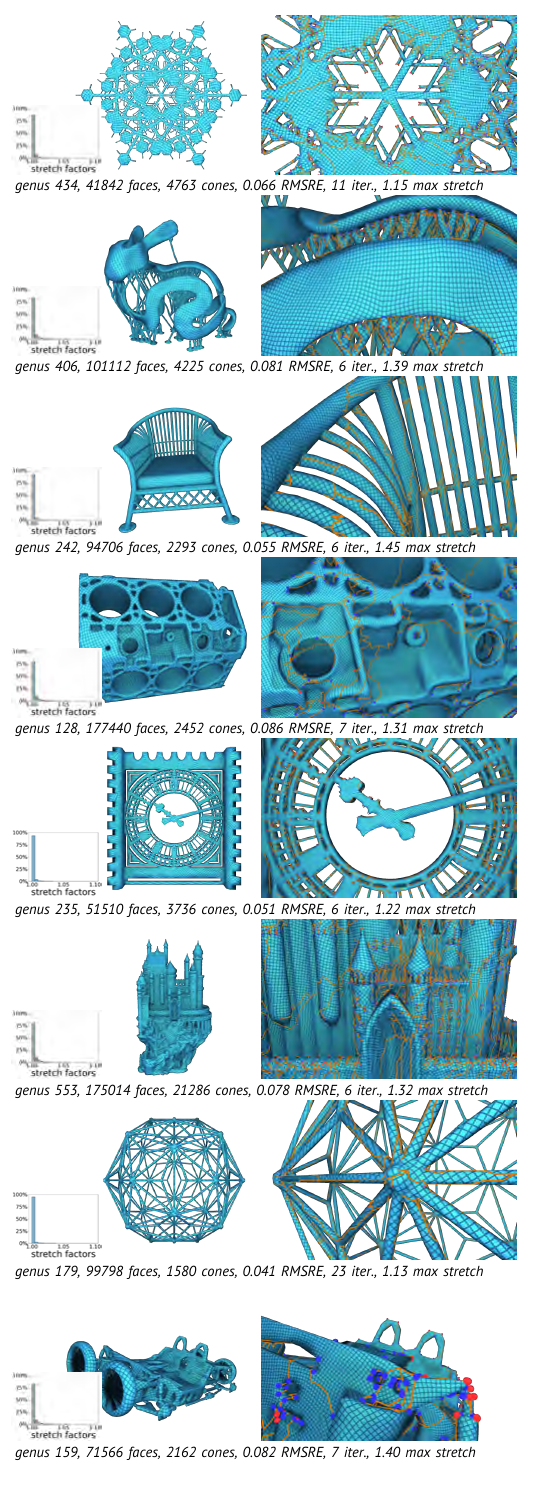}
     \caption{Parametrizations produced by our method for the modified Thingi10k dataset with per-edge symmetric stretch factors and mesh statistics.}
    \label{fig:tetwild-examples}
\end{figure*}

\paragraph{Topological robustness.}
To evaluate the reliability of our method on surface meshes with more varied and extreme topology, we also tested our method on a dataset derived from Thingi10k \citep{Thingi10K}. As our focus in this experiment is topological robustness, i.e., the ability of the algorithm to handle high genus joined with complex geometry, as well as complex fields, we used the remeshed version of the dataset included with \citet{hu2018tetrahedral}, which includes almost all meshes from the original dataset, but with triangle quality improved to have worst inradius to circumradius ratio typically above $10^{-4}$, in contrast to the original dataset containing many models with numerically degenerate triangles.  As a significant fraction of meshes produced by Tetwild are non-manifold, but not in a fundamentally difficult way (i.e., a union of manifold 3D domains attached at edges and/or vertices), we separated all non-manifold meshes into closed manifold components, and split all meshes with at most 10 components into separate meshes, and selected meshes with nontrivial genus. Smooth cross-fields and corresponding holonomy angle constraints were computed for this dataset, and meshes for which the resulting vertex angle constraints were theoretically unsatisfiable were discarded (3-5 torus topology \citep{shen2022cross}, 0 cone angles, and tori without cones but with nontrivial holonomy).  In total, we obtained 16156 meshes by this procedure. Statistics for this dataset and our method's performance on it are shown in Figure~\ref{fig:tetwild-dataset}. This dataset contains extremely high genus meshes that impose proportionately many loop holonomy angle constraints, but our method quickly produces parametrizations satisfying these constraints with low distortion. The output and performance of our method on some of the highest genus meshes in this dataset are presented in Figure~\ref{fig:tetwild-examples}.

\paragraph{Intrinsic preprocessing}

We evaluated both intrinsic preprocessing approaches, i.e., refinement and interpolation, on the closed manifold meshes of the original Thingi10k dataset \citep{Thingi10K}. After discarding meshes with degenerate triangles (with angles or lengths less than $10^{-10}$), we obtained a dataset of 5342 connected meshes with generally poor triangle quality. By splitting the disconnected meshes into separate manifold components, we obtained a larger dataset of 27180 meshes. In order to test on this full dataset, we applied simple heuristics (e.g., adding random cone pairs) to minimally modify any theoretically unsatisfiable constraints and produce a valid holonomy signature.

Unlike the extrinsic Tetwild remeshing, neither intrinsic method individually succeeded on the dataset for any choice of $\alpha_{\min}$. However, \emph{either} intrinsic refinement or interpolation succeeded on the \emph{full} dataset of 27180 meshes with some $\alpha_{\min}$ within 500 iterations. That is, we parameterize any closed nondegenerate mesh in the original Thingi10k dataset using only intrinsic methods for preprocessing.

\begin{figure}
    \centering
    \includegraphics[width=0.8\columnwidth]{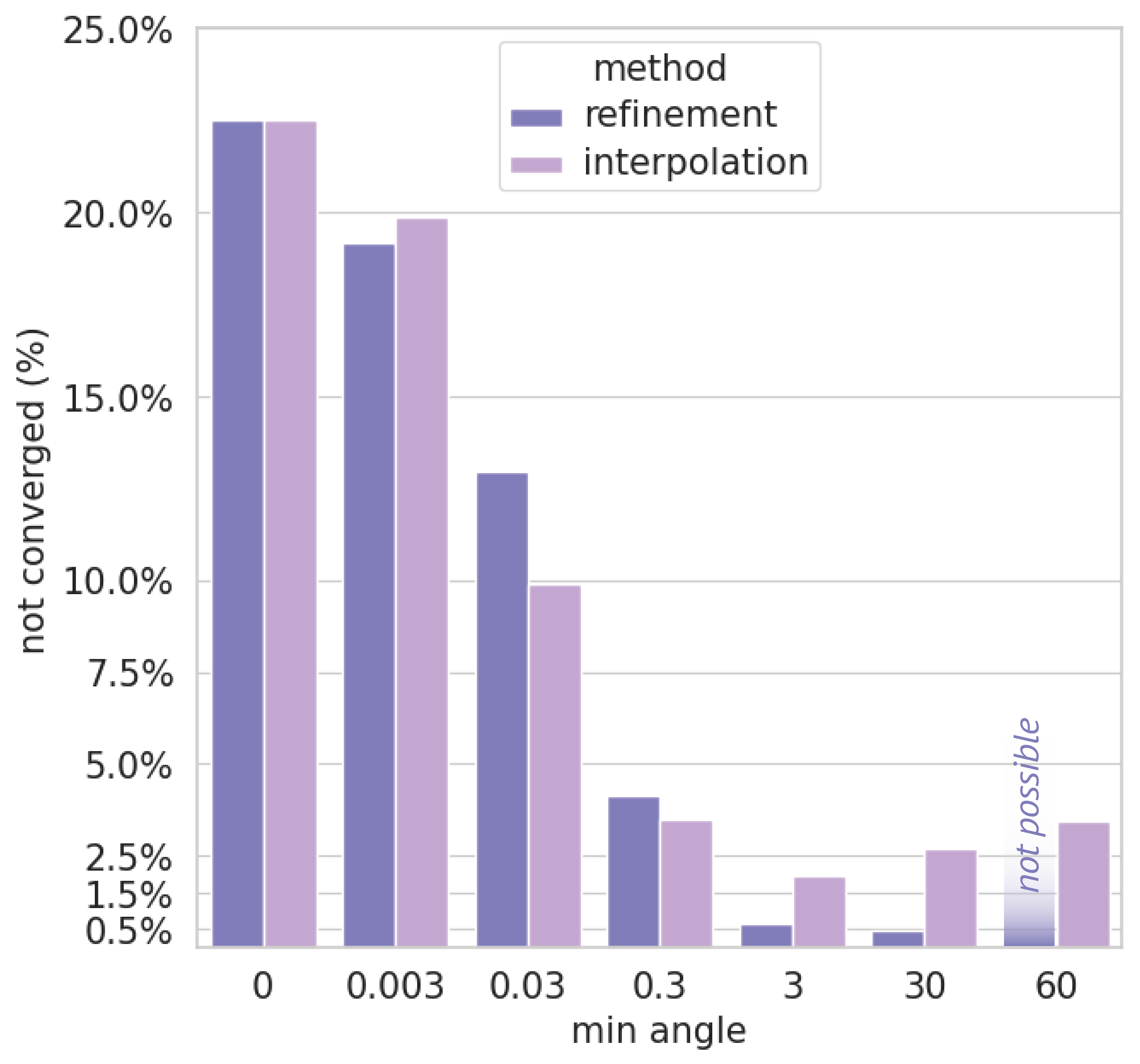}
    \caption{Convergence results on connected, nondegenerate meshes in the original Thingi10k dataset with intrinsic preprocessing. Refinement and interpolation are performed until some minimum triangle angle threshold $\alpha_{\min}$ is satisfied (away from needle-like features for refinement).}
    \label{fig:intrinsic}
\end{figure}

In Figure~\ref{fig:intrinsic}, we show the percentage of the 5342 connected models that failed to converge in 100 iterations with a timeout of 1 hour for increasing values of $\alpha_{\min}$. Note that for larger values of $\alpha_{\min}$ the number of failures with interpolation actually starts to increase. This degradation in performance is not surprising as the change in initial triangle angles $\alpha(\beta\lambda^0)$ resulting from small values of $\beta$ may increase the initial angle constraint error $F(\beta\lambda^0)$, e.g., at high valence vertices. More iterations may thus be required to satisfy the angle constraints, and poorly conditioned triangles can arise in later iterations. We also note that the intrinsic refinement increases the mesh size and thus increases the per-iteration cost of our method, although we found that the increase in per-iteration time is often well compensated by a decrease in the total number of iterations.

\paragraph{Metric optimality}

\begin{figure}
    \centering
    \includegraphics[width=\columnwidth]{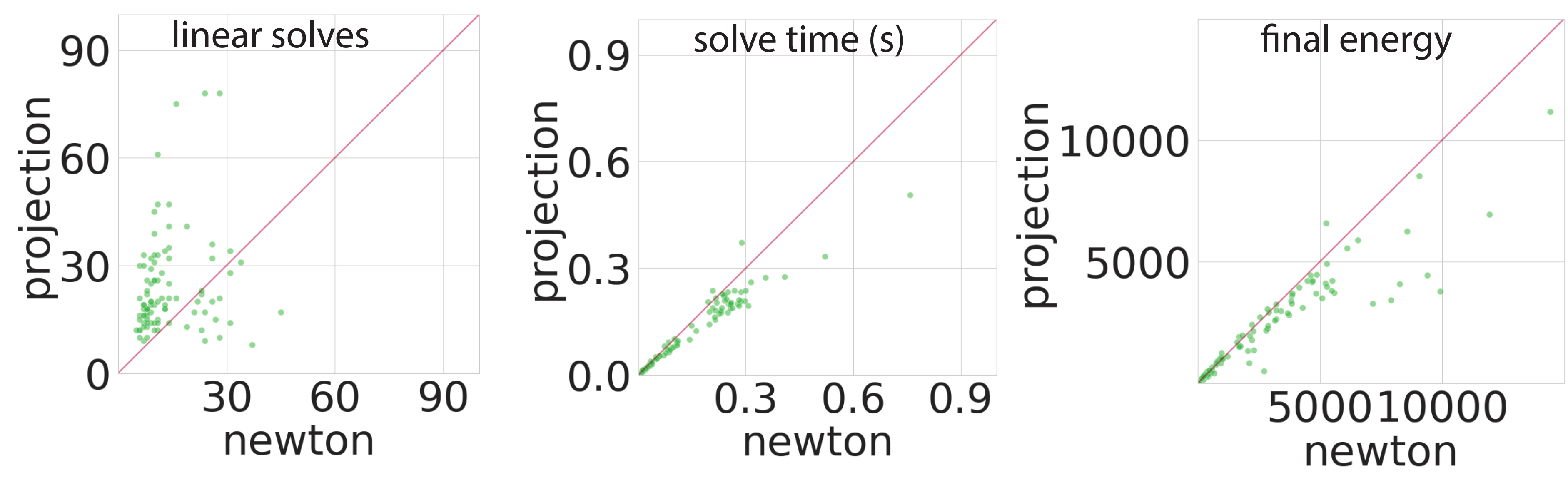}
    \caption{Comparison of the performance of our Newton method with the projected gradient descent method of \citet{capouellez2023metric} on the dataset from \citet{Myles:2014}. We plot (a) the number of linear solves required by our method to converge to $\max_\Theta \|\Theta - \hat\Theta\| \leq 10^{-12}$, against (b) the number of linear solves in projected gradient descent necessary to achieve the same geometric distortion $||\lambda - \lambda^0||_2$. We also compare the average time for each linear solve and the final energies after running the metric isometry optimization to convergence.}
    \label{fig:performance}
\end{figure}

As stated above, our algorithm minimizes the norm $||\lambda - \lambda^0||_2$ at each Newton iteration, but, since our constraints are nonlinear, our final solution only approximately minimizes this norm. In contrast, \citet{capouellez2023metric} explicitly optimizes the isometry energy $||\lambda - \lambda^0||_2^2$ while satisfying the same \emph{vertex} angle constraints as our method; however, this method provides no control over \emph{loop} holonomy angle constraints. In Figure~\ref{fig:performance}, we compare the performance of this alternative approach and our method \emph{with only vertex angle constraints}. Our method produces a comparable amount of geometric distortion while on average requiring fewer linear solves.

\begin{figure*}[h]
    \centering
    \includegraphics[width=0.85\textwidth]{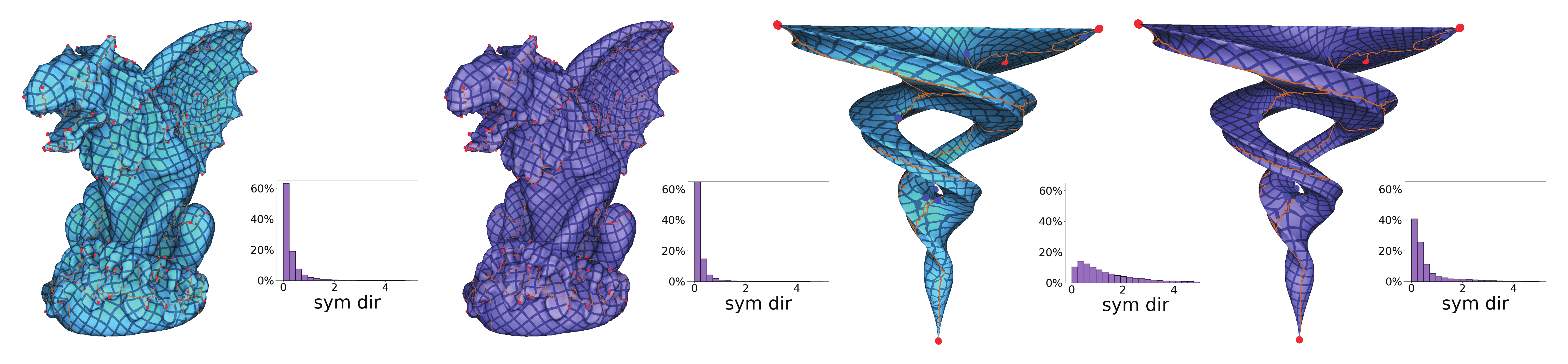}
    \caption{Comparison of quality of parametrizations before (blue) and after (purple) symmetric Dirichlet optimization with seamless constraints. Histograms show the distribution of per face symmetric Dirichlet values. Left: our parameterization already has low geometric distortion, and is largely unchanged by further optimization. Right: intrinsic metric interpolation was used to improve the initial triangle quality, and our parameterization is distorted, but symmetric Dirichlet optimization is able to produce a visually pleasing result.}
    \label{fig:quality}
\end{figure*}

\paragraph{Symmetric Dirichlet post-processing}
Once an initial seamless parameterization is obtained, it can be robustly optimized further to improve isometry using the symmetric Dirichlet energy by Newton's method with linear constraints to preserve seamlessness. We optimized our parameterizations of the dataset from \citet{Myles:2014}; examples of parameterizations before and after optimization are demonstrated in Figure~\ref{fig:quality}. Unlike conformal methods, which can produce extreme geometric distortion and consequently numerical instability, our method produces initial parameterizations that are already approximately isometric and thus amenable to such optimization, and in most cases such post-processing is not needed.

\begin{figure*}[h]
    \centering
    \includegraphics[width=0.85 \textwidth]{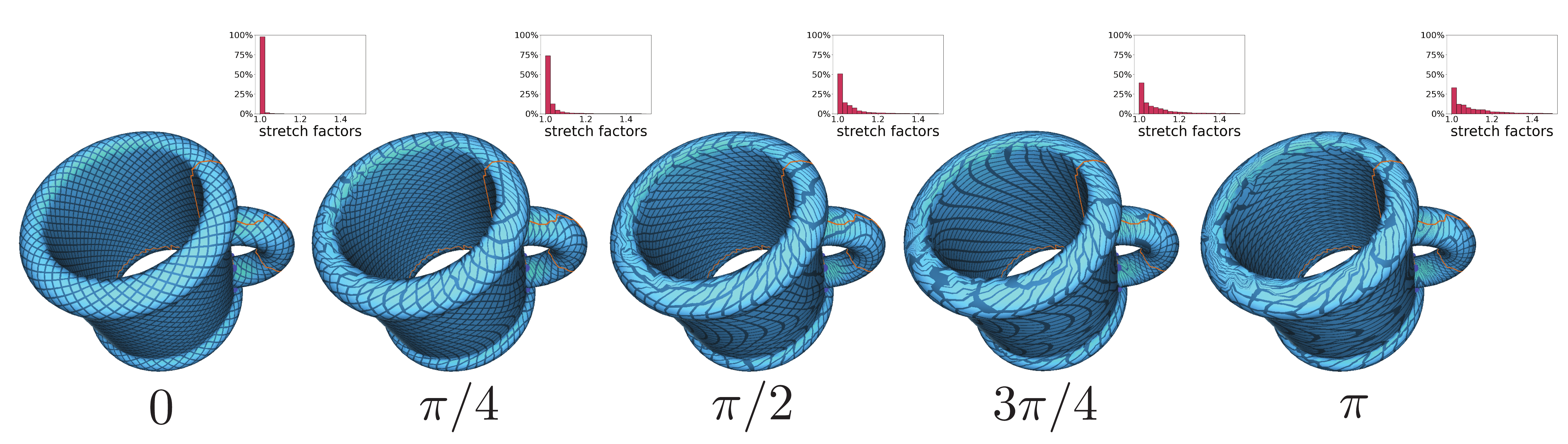}
        \caption{Parametrizations produced by our
        method with an increasing holonomy angle constraint for a single dual loop. We increase the angle constraint from an initial value arising from a cross field in increments of $\pi/4$.}
        \label{fig:stress-test}
    \end{figure*}

\paragraph{Arbitrary loop holonomy angle constraints}
While we primarily focus on parameterizations satisfying angle constraints arising from frame or cross-fields, our method also supports arbitrary holonomy angle constraints, including angle constraints that are not integral multiples of $\pi/2$. We demonstrate the effect of increasing the constraint for a single dual loop on the resulting parameterization in Figure~\ref{fig:stress-test}. While the parameterization quality suffers from this extreme geometry distorting constraint, our method is still able to robustly produce a valid solution.

\paragraph{Robustness compared to state-of-the-art $UV$ and hybrid methods.} 
Our method succeeded on the closed meshes of  the dataset introduced in \citet{Myles:2014} and used in a few papers, with original connectivity and fields as input, as well as on 100\% of 16,147 mesh dataset described above.
We focus on methods that solve the seamless parametrization problem, i.e., aim to satisfy both angle and loop constraints.

The method of \citet{Myles:2014}  succeeds in producing a seamless parametrization (along with T-mesh partition) for all shapes, but for four, all closed, needs to add cones, and for 39 required T-mesh modification, which may result in changes in holonomy on loops.  Three other methods evaluated in that paper are: original MIQ \citep{Bommes:2009}, the IGM method \citep{Bommes:2013} (did not find a solution in 25\%  of cases), and MIQ combined with the convexified bijectivity constraints \citep{Lipman:2012} (did not find a solution in 17\%). Analyzing the data, we also observe that while the remaining parameterizations are always bijective, they often introduce integer-index cones, i.e., do not preserve holonomy signature.

More recent work includes \citet{Campen:2017:OnSimilarityMaps} which succeeds on the whole dataset of \citet{Myles:2014}, but produces only similarity maps, not seamless maps. Most importantly, the recent method  \citet{shen2022cross}, while having theoretical guarantees, fails on the highest genus models (6\% of nontrivial genus models, highest genus around 100)  in \citet{Myles:2014} due to the extreme distortion of the intermediate maps it generates, and requiring expensive path rerouting and optimization steps in the end. In contrast, our method succeeds on the whole set (to the best of our knowledge, the first one, without using cone insertion or cone or holonomy angle modification).

\section{Concluding Remarks}

The main limitation of our algorithm that we have observed so far is sensitivity to extremely bad mesh quality, which is somewhat higher than for conformal maps. 
At the same time, its performance on a large dataset, along with theoretical considerations suggests that there are fundamental reasons  
for its convergence behavior, in particular, that can be connected to convex function optimization.  

We identify several promising directions for future research: 
(1) placing the algorithm on a solid theoretical basis, and fully describing constraints on the holonomy signature for which it does not produce a seamless parametrization. 
(2) While the extension to handling boundaries is straightforward, following the doubling process described in \citet{campen2021efficient}, the extension to constraints on sharp features (e.g., requiring features to be straight and axis aligned in parametric domain) requires more work.  Techniques for doing this in angle variables were developed in \citet{myles2013controlled}.
(3) The algorithm produces feasible seamless parametrizations but optimizes a specific energy only indirectly.
(4) Due to its simplicity, it is amenable to domain decomposition and hierarchical extensions, to scale it to large meshes.

\newpage
\clearpage
\bibliographystyle{ACM-Reference-Format}
\bibliography{99-bib}


\begin{thebibliography}{46}


\ifx \showCODEN    \undefined \def \showCODEN     #1{\unskip}     \fi
\ifx \showDOI      \undefined \def \showDOI       #1{#1}\fi
\ifx \showISBNx    \undefined \def \showISBNx     #1{\unskip}     \fi
\ifx \showISBNxiii \undefined \def \showISBNxiii  #1{\unskip}     \fi
\ifx \showISSN     \undefined \def \showISSN      #1{\unskip}     \fi
\ifx \showLCCN     \undefined \def \showLCCN      #1{\unskip}     \fi
\ifx \shownote     \undefined \def \shownote      #1{#1}          \fi
\ifx \showarticletitle \undefined \def \showarticletitle #1{#1}   \fi
\ifx \showURL      \undefined \def \showURL       {\relax}        \fi
\providecommand\bibfield[2]{#2}
\providecommand\bibinfo[2]{#2}
\providecommand\natexlab[1]{#1}
\providecommand\showeprint[2][]{arXiv:#2}

\bibitem[Ben-Chen et~al\mbox{.}(2008)]%
        {BenChen:2008}
\bibfield{author}{\bibinfo{person}{Mirela Ben-Chen}, \bibinfo{person}{Craig Gotsman}, {and} \bibinfo{person}{Guy Bunin}.} \bibinfo{year}{2008}\natexlab{}.
\newblock \showarticletitle{{Conformal Flattening by Curvature Prescription and Metric Scaling}}.
\newblock \bibinfo{journal}{\emph{Computer Graphics Forum}} \bibinfo{volume}{27}, \bibinfo{number}{2} (\bibinfo{year}{2008}).
\newblock


\bibitem[Bommes et~al\mbox{.}(2013)]%
        {Bommes:2013}
\bibfield{author}{\bibinfo{person}{David Bommes}, \bibinfo{person}{Marcel Campen}, \bibinfo{person}{Hans-Christian Ebke}, \bibinfo{person}{Pierre Alliez}, {and} \bibinfo{person}{Leif Kobbelt}.} \bibinfo{year}{2013}\natexlab{}.
\newblock \showarticletitle{Integer-Grid Maps for Reliable Quad Meshing}.
\newblock \bibinfo{journal}{\emph{ACM Trans. Graph.}} \bibinfo{volume}{32}, \bibinfo{number}{4} (\bibinfo{year}{2013}), \bibinfo{pages}{98:1--98:12}.
\newblock


\bibitem[Bommes et~al\mbox{.}(2009)]%
        {Bommes:2009}
\bibfield{author}{\bibinfo{person}{David Bommes}, \bibinfo{person}{Henrik Zimmer}, {and} \bibinfo{person}{Leif Kobbelt}.} \bibinfo{year}{2009}\natexlab{}.
\newblock \showarticletitle{Mixed-integer quadrangulation}.
\newblock \bibinfo{journal}{\emph{ACM Trans. Graph.}} \bibinfo{volume}{28}, \bibinfo{number}{3} (\bibinfo{year}{2009}), \bibinfo{pages}{77}.
\newblock


\bibitem[Bright et~al\mbox{.}(2017)]%
        {Bright:2017:HGP}
\bibfield{author}{\bibinfo{person}{Alon Bright}, \bibinfo{person}{Edward Chien}, {and} \bibinfo{person}{Ofir Weber}.} \bibinfo{year}{2017}\natexlab{}.
\newblock \showarticletitle{Harmonic Global Parametrization with Rational Holonomy}.
\newblock \bibinfo{journal}{\emph{$\!$ACM Trans.$\!$ Graph.}} \bibinfo{volume}{$\!$36}, \bibinfo{number}{$\!\,$4} (\bibinfo{year}{2017}).
\newblock


\bibitem[Campen et~al\mbox{.}(2015)]%
        {CampenBK15}
\bibfield{author}{\bibinfo{person}{Marcel Campen}, \bibinfo{person}{David Bommes}, {and} \bibinfo{person}{Leif Kobbelt}.} \bibinfo{year}{2015}\natexlab{}.
\newblock \showarticletitle{Quantized global parametrization}.
\newblock \bibinfo{journal}{\emph{ACM Trans. Graph.}} \bibinfo{volume}{34}, \bibinfo{number}{6} (\bibinfo{year}{2015}), \bibinfo{pages}{192}.
\newblock


\bibitem[Campen et~al\mbox{.}(2021)]%
        {campen2021efficient}
\bibfield{author}{\bibinfo{person}{Marcel Campen}, \bibinfo{person}{Ryan Capouellez}, \bibinfo{person}{Hanxiao Shen}, \bibinfo{person}{Leyi Zhu}, \bibinfo{person}{Daniele Panozzo}, {and} \bibinfo{person}{Denis Zorin}.} \bibinfo{year}{2021}\natexlab{}.
\newblock \showarticletitle{Efficient and Robust Discrete Conformal Equivalence with Boundary}.
\newblock \bibinfo{journal}{\emph{ACM Trans. Graph.}} \bibinfo{volume}{40}, \bibinfo{number}{6}, Article \bibinfo{articleno}{261} (\bibinfo{date}{dec} \bibinfo{year}{2021}), \bibinfo{numpages}{16}~pages.
\newblock
\showISSN{0730-0301}
\urldef\tempurl%
\url{https://doi.org/10.1145/3478513.3480557}
\showDOI{\tempurl}


\bibitem[Campen et~al\mbox{.}(2019)]%
        {campen2018seamless}
\bibfield{author}{\bibinfo{person}{Marcel Campen}, \bibinfo{person}{Hanxiao Shen}, \bibinfo{person}{Jiaran Zhou}, {and} \bibinfo{person}{Denis Zorin}.} \bibinfo{year}{2019}\natexlab{}.
\newblock \showarticletitle{Seamless Parametrization with Arbitrary Cones for Arbitrary Genus}.
\newblock \bibinfo{journal}{\emph{ACM Trans. Graph.}} \bibinfo{volume}{39}, \bibinfo{number}{1} (\bibinfo{year}{2019}).
\newblock


\bibitem[Campen and Zorin(2017a)]%
        {Campen:2017:OnSimilarityMaps}
\bibfield{author}{\bibinfo{person}{Marcel Campen} {and} \bibinfo{person}{Denis Zorin}.} \bibinfo{year}{2017}\natexlab{a}.
\newblock \bibinfo{booktitle}{\emph{On Discrete Conformal Seamless Similarity Maps}}.
\newblock
\showeprint[arxiv]{1705.02422}~[cs.GR]


\bibitem[Campen and Zorin(2017b)]%
        {Campen:2017:SimilarityMaps}
\bibfield{author}{\bibinfo{person}{Marcel Campen} {and} \bibinfo{person}{Denis Zorin}.} \bibinfo{year}{2017}\natexlab{b}.
\newblock \showarticletitle{Similarity Maps and Field-Guided T-Splines: a Perfect Couple}.
\newblock \bibinfo{journal}{\emph{ACM Trans. Graph.}} \bibinfo{volume}{36}, \bibinfo{number}{4} (\bibinfo{year}{2017}).
\newblock


\bibitem[Capouellez and Zorin(2023)]%
        {capouellez2023metric}
\bibfield{author}{\bibinfo{person}{Ryan Capouellez} {and} \bibinfo{person}{Denis Zorin}.} \bibinfo{year}{2023}\natexlab{}.
\newblock \showarticletitle{Metric Optimization in Penner Coordinates}.
\newblock \bibinfo{journal}{\emph{ACM Trans. Graph.}} \bibinfo{volume}{42}, \bibinfo{number}{6}, Article \bibinfo{articleno}{234} (\bibinfo{date}{dec} \bibinfo{year}{2023}), \bibinfo{numpages}{19}~pages.
\newblock
\showISSN{0730-0301}
\urldef\tempurl%
\url{https://doi.org/10.1145/3618394}
\showDOI{\tempurl}


\bibitem[Crane et~al\mbox{.}(2010)]%
        {CraneTrivial}
\bibfield{author}{\bibinfo{person}{Keenan Crane}, \bibinfo{person}{Mathieu Desbrun}, {and} \bibinfo{person}{Peter Schröder}.} \bibinfo{year}{2010}\natexlab{}.
\newblock \showarticletitle{Trivial Connections on Discrete Surfaces}.
\newblock \bibinfo{journal}{\emph{Computer Graphics Forum}} \bibinfo{volume}{29}, \bibinfo{number}{5} (\bibinfo{year}{2010}), \bibinfo{pages}{1525--1533}.
\newblock


\bibitem[Farchi and Ben-Chen(2018)]%
        {farchi2018integer}
\bibfield{author}{\bibinfo{person}{Nahum Farchi} {and} \bibinfo{person}{Mirela Ben-Chen}.} \bibinfo{year}{2018}\natexlab{}.
\newblock \showarticletitle{Integer-only cross field computation}.
\newblock \bibinfo{journal}{\emph{ACM Trans. Graph.}} \bibinfo{volume}{37}, \bibinfo{number}{4}, Article \bibinfo{articleno}{91} (\bibinfo{date}{jul} \bibinfo{year}{2018}), \bibinfo{numpages}{13}~pages.
\newblock
\showISSN{0730-0301}
\urldef\tempurl%
\url{https://doi.org/10.1145/3197517.3201375}
\showDOI{\tempurl}


\bibitem[Fu et~al\mbox{.}(2021)]%
        {fu2021inversion}
\bibfield{author}{\bibinfo{person}{Xiao-Ming Fu}, \bibinfo{person}{Jian-Ping Su}, \bibinfo{person}{Zheng-Yu Zhao}, \bibinfo{person}{Qing Fang}, \bibinfo{person}{Chunyang Ye}, {and} \bibinfo{person}{Ligang Liu}.} \bibinfo{year}{2021}\natexlab{}.
\newblock \showarticletitle{Inversion-free geometric mapping construction: A survey}.
\newblock \bibinfo{journal}{\emph{Computational Visual Media}} \bibinfo{volume}{7}, \bibinfo{number}{3} (\bibinfo{year}{2021}), \bibinfo{pages}{289--318}.
\newblock


\bibitem[Gillespie et~al\mbox{.}(2021)]%
        {gillespie2021discrete}
\bibfield{author}{\bibinfo{person}{Mark Gillespie}, \bibinfo{person}{Boris Springborn}, {and} \bibinfo{person}{Keenan Crane}.} \bibinfo{year}{2021}\natexlab{}.
\newblock \showarticletitle{Discrete Conformal Equivalence of Polyhedral Surfaces}.
\newblock \bibinfo{journal}{\emph{ACM Trans. Graph.}} \bibinfo{volume}{40}, \bibinfo{number}{4}, Article \bibinfo{articleno}{103} (\bibinfo{date}{jul} \bibinfo{year}{2021}), \bibinfo{numpages}{20}~pages.
\newblock
\showISSN{0730-0301}
\urldef\tempurl%
\url{https://doi.org/10.1145/3450626.3459763}
\showDOI{\tempurl}


\bibitem[Gu et~al\mbox{.}(2018a)]%
        {gu2018discrete2}
\bibfield{author}{\bibinfo{person}{Xianfeng Gu}, \bibinfo{person}{Ren Guo}, \bibinfo{person}{Feng Luo}, \bibinfo{person}{Jian Sun}, {and} \bibinfo{person}{Tianqi Wu}.} \bibinfo{year}{2018}\natexlab{a}.
\newblock \showarticletitle{A discrete uniformization theorem for polyhedral surfaces II}.
\newblock \bibinfo{journal}{\emph{Journal of Differential Geometry}} \bibinfo{volume}{109}, \bibinfo{number}{3} (\bibinfo{year}{2018}), \bibinfo{pages}{431--466}.
\newblock


\bibitem[Gu et~al\mbox{.}(2018b)]%
        {gu2018discrete}
\bibfield{author}{\bibinfo{person}{Xianfeng Gu}, \bibinfo{person}{Feng Luo}, \bibinfo{person}{Jian Sun}, {and} \bibinfo{person}{Tianqi Wu}.} \bibinfo{year}{2018}\natexlab{b}.
\newblock \showarticletitle{A discrete uniformization theorem for polyhedral surfaces}.
\newblock \bibinfo{journal}{\emph{Journal of Differential Geometry}} \bibinfo{volume}{109}, \bibinfo{number}{2} (\bibinfo{year}{2018}), \bibinfo{pages}{223--256}.
\newblock


\bibitem[Hefetz et~al\mbox{.}(2019)]%
        {hefetz2019subspace}
\bibfield{author}{\bibinfo{person}{Eden~Fedida Hefetz}, \bibinfo{person}{Edward Chien}, {and} \bibinfo{person}{Ofir Weber}.} \bibinfo{year}{2019}\natexlab{}.
\newblock \showarticletitle{A Subspace Method for Fast Locally Injective Harmonic Mapping}.
\newblock \bibinfo{journal}{\emph{Computer Graphics Forum}} \bibinfo{volume}{38}, \bibinfo{number}{2} (\bibinfo{year}{2019}), \bibinfo{pages}{105--119}.
\newblock


\bibitem[Hu et~al\mbox{.}(2018)]%
        {hu2018tetrahedral}
\bibfield{author}{\bibinfo{person}{Yixin Hu}, \bibinfo{person}{Qingnan Zhou}, \bibinfo{person}{Xifeng Gao}, \bibinfo{person}{Alec Jacobson}, \bibinfo{person}{Denis Zorin}, {and} \bibinfo{person}{Daniele Panozzo}.} \bibinfo{year}{2018}\natexlab{}.
\newblock \showarticletitle{Tetrahedral meshing in the wild.}
\newblock \bibinfo{journal}{\emph{ACM Trans. Graph.}} \bibinfo{volume}{37}, \bibinfo{number}{4} (\bibinfo{year}{2018}), \bibinfo{pages}{60--1}.
\newblock


\bibitem[Izmestiev et~al\mbox{.}(2013)]%
        {izmestiev2013there}
\bibfield{author}{\bibinfo{person}{Ivan Izmestiev}, \bibinfo{person}{Robert~B Kusner}, \bibinfo{person}{G{\"u}nter Rote}, \bibinfo{person}{Boris Springborn}, {and} \bibinfo{person}{John~M Sullivan}.} \bibinfo{year}{2013}\natexlab{}.
\newblock \showarticletitle{There is no triangulation of the torus with vertex degrees 5, 6,..., 6, 7 and related results: Geometric proofs for combinatorial theorems}.
\newblock \bibinfo{journal}{\emph{Geometriae Dedicata}}  \bibinfo{volume}{166} (\bibinfo{year}{2013}), \bibinfo{pages}{15--29}.
\newblock


\bibitem[Jucovi{\v{c}} and Trenkler(1973)]%
        {jucovivc1973theorem}
\bibfield{author}{\bibinfo{person}{Ernest Jucovi{\v{c}}} {and} \bibinfo{person}{Mari\'{a}n Trenkler}.} \bibinfo{year}{1973}\natexlab{}.
\newblock \showarticletitle{A theorem on the structure of cell--decompositions of orientable 2--manifolds}.
\newblock \bibinfo{journal}{\emph{Mathematika}} \bibinfo{volume}{20}, \bibinfo{number}{01} (\bibinfo{year}{1973}), \bibinfo{pages}{63--82}.
\newblock


\bibitem[K{\"a}lberer et~al\mbox{.}(2007)]%
        {kalberer2007qsp}
\bibfield{author}{\bibinfo{person}{F. K{\"a}lberer}, \bibinfo{person}{M. Nieser}, {and} \bibinfo{person}{K. Polthier}.} \bibinfo{year}{2007}\natexlab{}.
\newblock \showarticletitle{{QuadCover: Surface Parameterization using Branched Coverings}}.
\newblock \bibinfo{journal}{\emph{Computer Graphics Forum}} \bibinfo{volume}{26}, \bibinfo{number}{3} (\bibinfo{year}{2007}), \bibinfo{pages}{375--384}.
\newblock


\bibitem[Kharevych et~al\mbox{.}(2006)]%
        {kharevych2006discrete}
\bibfield{author}{\bibinfo{person}{Liliya Kharevych}, \bibinfo{person}{Boris Springborn}, {and} \bibinfo{person}{Peter Schr{\"o}der}.} \bibinfo{year}{2006}\natexlab{}.
\newblock \showarticletitle{Discrete conformal mappings via circle patterns}.
\newblock \bibinfo{journal}{\emph{ACM Trans. Graph.}} \bibinfo{volume}{25}, \bibinfo{number}{2} (\bibinfo{year}{2006}), \bibinfo{pages}{412--438}.
\newblock


\bibitem[Levi(2023)]%
        {levi2023seamless}
\bibfield{author}{\bibinfo{person}{Zohar Levi}.} \bibinfo{year}{2023}\natexlab{}.
\newblock \showarticletitle{Seamless Parametrization with Cone and Partial Loop Control}.
\newblock \bibinfo{journal}{\emph{ACM Transactions on Graphics}} (\bibinfo{year}{2023}).
\newblock


\bibitem[{Li} et~al\mbox{.}(2006)]%
        {LiPeriodJumps}
\bibfield{author}{\bibinfo{person}{W. {Li}}, \bibinfo{person}{B. {Vallet}}, \bibinfo{person}{N. {Ray}}, {and} \bibinfo{person}{B. {Levy}}.} \bibinfo{year}{2006}\natexlab{}.
\newblock \showarticletitle{Representing Higher-Order Singularities in Vector Fields on Piecewise Linear Surfaces}.
\newblock \bibinfo{journal}{\emph{IEEE Transactions on Visualization and Computer Graphics}} \bibinfo{volume}{12}, \bibinfo{number}{5} (\bibinfo{year}{2006}), \bibinfo{pages}{1315--1322}.
\newblock


\bibitem[Lipman(2012)]%
        {Lipman:2012}
\bibfield{author}{\bibinfo{person}{Yaron Lipman}.} \bibinfo{year}{2012}\natexlab{}.
\newblock \showarticletitle{Bounded Distortion Mapping Spaces for Triangular Meshes}.
\newblock \bibinfo{journal}{\emph{ACM Trans. Graph.}} \bibinfo{volume}{31}, \bibinfo{number}{4} (\bibinfo{year}{2012}), \bibinfo{pages}{108:1--108:13}.
\newblock


\bibitem[Liu et~al\mbox{.}(2018)]%
        {liu2018progressive}
\bibfield{author}{\bibinfo{person}{Ligang Liu}, \bibinfo{person}{Chunyang Ye}, \bibinfo{person}{Ruiqi Ni}, {and} \bibinfo{person}{Xiao-Ming Fu}.} \bibinfo{year}{2018}\natexlab{}.
\newblock \showarticletitle{Progressive parameterizations}.
\newblock \bibinfo{journal}{\emph{ACM Transactions on Graphics (TOG)}} \bibinfo{volume}{37}, \bibinfo{number}{4} (\bibinfo{year}{2018}), \bibinfo{pages}{1--12}.
\newblock


\bibitem[Lyon et~al\mbox{.}(2021)]%
        {lyon2021quad}
\bibfield{author}{\bibinfo{person}{Max Lyon}, \bibinfo{person}{Marcel Campen}, {and} \bibinfo{person}{Leif Kobbelt}.} \bibinfo{year}{2021}\natexlab{}.
\newblock \showarticletitle{Quad layouts via constrained t-mesh quantization}. In \bibinfo{booktitle}{\emph{Computer Graphics Forum}}, Vol.~\bibinfo{volume}{40}. Wiley Online Library, \bibinfo{pages}{305--314}.
\newblock


\bibitem[Myles et~al\mbox{.}(2014)]%
        {Myles:2014}
\bibfield{author}{\bibinfo{person}{Ashish Myles}, \bibinfo{person}{Nico Pietroni}, {and} \bibinfo{person}{Denis Zorin}.} \bibinfo{year}{2014}\natexlab{}.
\newblock \showarticletitle{Robust Field-aligned Global Parametrization}.
\newblock \bibinfo{journal}{\emph{ACM Trans. Graph.}} \bibinfo{volume}{33}, \bibinfo{number}{4} (\bibinfo{year}{2014}), \bibinfo{pages}{135:1--135:14}.
\newblock


\bibitem[Myles and Zorin(2013)]%
        {myles2013controlled}
\bibfield{author}{\bibinfo{person}{Ashish Myles} {and} \bibinfo{person}{Denis Zorin}.} \bibinfo{year}{2013}\natexlab{}.
\newblock \showarticletitle{Controlled-distortion constrained global parametrization}.
\newblock \bibinfo{journal}{\emph{ACM Transactions on Graphics}} \bibinfo{volume}{32}, \bibinfo{number}{4} (\bibinfo{year}{2013}), \bibinfo{pages}{105}.
\newblock


\bibitem[Naitsat et~al\mbox{.}(2021)]%
        {naitsat2021inversion}
\bibfield{author}{\bibinfo{person}{Alexander Naitsat}, \bibinfo{person}{Gregory Naitzat}, {and} \bibinfo{person}{Yehoshua~Y Zeevi}.} \bibinfo{year}{2021}\natexlab{}.
\newblock \showarticletitle{On Inversion-Free Mapping and Distortion Minimization}.
\newblock \bibinfo{journal}{\emph{Journal of Mathematical Imaging and Vision}} (\bibinfo{year}{2021}), \bibinfo{pages}{1--36}.
\newblock


\bibitem[Penner(1987)]%
        {penner1987decorated}
\bibfield{author}{\bibinfo{person}{Robert~C Penner}.} \bibinfo{year}{1987}\natexlab{}.
\newblock \showarticletitle{The decorated Teichm{\"u}ller space of punctured surfaces}.
\newblock \bibinfo{journal}{\emph{Communications in Mathematical Physics}} \bibinfo{volume}{113}, \bibinfo{number}{2} (\bibinfo{year}{1987}), \bibinfo{pages}{299--339}.
\newblock


\bibitem[Rabinovich et~al\mbox{.}(2017)]%
        {Rabinovich:2017:SLI}
\bibfield{author}{\bibinfo{person}{Michael Rabinovich}, \bibinfo{person}{Roi Poranne}, \bibinfo{person}{Daniele Panozzo}, {and} \bibinfo{person}{Olga Sorkine-Hornung}.} \bibinfo{year}{2017}\natexlab{}.
\newblock \showarticletitle{Scalable Locally Injective Mappings}.
\newblock \bibinfo{journal}{\emph{ACM Trans. Graph.}} \bibinfo{volume}{36}, \bibinfo{number}{2} (\bibinfo{year}{2017}), \bibinfo{pages}{16:1--16:16}.
\newblock


\bibitem[Ray et~al\mbox{.}(2009)]%
        {Ray:2009:GeometryAware}
\bibfield{author}{\bibinfo{person}{Nicolas Ray}, \bibinfo{person}{Bruno Vallet}, \bibinfo{person}{Laurent Alonso}, {and} \bibinfo{person}{Bruno Levy}.} \bibinfo{year}{2009}\natexlab{}.
\newblock \showarticletitle{Geometry-Aware Direction Field Processing}.
\newblock \bibinfo{journal}{\emph{ACM Trans. Graph.}} \bibinfo{volume}{29}, \bibinfo{number}{1} (\bibinfo{year}{2009}).
\newblock


\bibitem[Ray et~al\mbox{.}(2008)]%
        {NSymmetry}
\bibfield{author}{\bibinfo{person}{Nicolas Ray}, \bibinfo{person}{Bruno Vallet}, \bibinfo{person}{Wan~Chiu Li}, {and} \bibinfo{person}{Bruno L\'{e}vy}.} \bibinfo{year}{2008}\natexlab{}.
\newblock \showarticletitle{N-Symmetry Direction Field Design}.
\newblock \bibinfo{journal}{\emph{ACM Trans. Graph.}} \bibinfo{volume}{27}, \bibinfo{number}{2} (\bibinfo{year}{2008}).
\newblock


\bibitem[Rivin(1994)]%
        {rivin1994euclidean}
\bibfield{author}{\bibinfo{person}{Igor Rivin}.} \bibinfo{year}{1994}\natexlab{}.
\newblock \showarticletitle{Euclidean structures on simplicial surfaces and hyperbolic volume}.
\newblock \bibinfo{journal}{\emph{Annals of mathematics}} \bibinfo{volume}{139}, \bibinfo{number}{3} (\bibinfo{year}{1994}), \bibinfo{pages}{553--580}.
\newblock


\bibitem[Sch{\"u}ller et~al\mbox{.}(2013)]%
        {Schueller:LIM:2013}
\bibfield{author}{\bibinfo{person}{Christian Sch{\"u}ller}, \bibinfo{person}{Ladislav Kavan}, \bibinfo{person}{Daniele Panozzo}, {and} \bibinfo{person}{Olga Sorkine-Hornung}.} \bibinfo{year}{2013}\natexlab{}.
\newblock \showarticletitle{Locally Injective Mappings}.
\newblock \bibinfo{journal}{\emph{Computer Graphics Forum}} \bibinfo{volume}{32}, \bibinfo{number}{5} (\bibinfo{year}{2013}), \bibinfo{pages}{125--135}.
\newblock


\bibitem[Sharp et~al\mbox{.}(2019a)]%
        {geometrycentral}
\bibfield{author}{\bibinfo{person}{Nicholas Sharp}, \bibinfo{person}{Keenan Crane}, {et~al\mbox{.}}} \bibinfo{year}{2019}\natexlab{a}.
\newblock \showarticletitle{GeometryCentral: A modern C++ library of data structures and algorithms for geometry processing}.
\newblock \bibinfo{howpublished}{\url{https://geometry-central.net/}}.
\newblock  (\bibinfo{year}{2019}).
\newblock


\bibitem[Sharp et~al\mbox{.}(2019b)]%
        {sharp2019navigating}
\bibfield{author}{\bibinfo{person}{Nicholas Sharp}, \bibinfo{person}{Yousuf Soliman}, {and} \bibinfo{person}{Keenan Crane}.} \bibinfo{year}{2019}\natexlab{b}.
\newblock \showarticletitle{Navigating intrinsic triangulations}.
\newblock \bibinfo{journal}{\emph{ACM Transactions on Graphics}} \bibinfo{volume}{38}, \bibinfo{number}{4} (\bibinfo{year}{2019}), \bibinfo{pages}{1--16}.
\newblock


\bibitem[Sheffer and de~Sturler(2001)]%
        {sheffer2001parameterization}
\bibfield{author}{\bibinfo{person}{Alla Sheffer} {and} \bibinfo{person}{Eric de Sturler}.} \bibinfo{year}{2001}\natexlab{}.
\newblock \showarticletitle{Parameterization of faceted surfaces for meshing using angle-based flattening}.
\newblock \bibinfo{journal}{\emph{Engineering with computers}} \bibinfo{volume}{17}, \bibinfo{number}{3} (\bibinfo{year}{2001}), \bibinfo{pages}{326--337}.
\newblock


\bibitem[Shen et~al\mbox{.}(2022)]%
        {shen2022cross}
\bibfield{author}{\bibinfo{person}{Hanxiao Shen}, \bibinfo{person}{Leyi Zhu}, \bibinfo{person}{Ryan Capouellez}, \bibinfo{person}{Daniele Panozzo}, \bibinfo{person}{Marcel Campen}, {and} \bibinfo{person}{Denis Zorin}.} \bibinfo{year}{2022}\natexlab{}.
\newblock \showarticletitle{Which cross fields can be quadrangulated? global parameterization from prescribed holonomy signatures}.
\newblock \bibinfo{journal}{\emph{ACM Transactions on Graphics (TOG)}} \bibinfo{volume}{41}, \bibinfo{number}{4} (\bibinfo{year}{2022}), \bibinfo{pages}{1--12}.
\newblock


\bibitem[Springborn(2020)]%
        {springborn2019ideal}
\bibfield{author}{\bibinfo{person}{Boris Springborn}.} \bibinfo{year}{2020}\natexlab{}.
\newblock \showarticletitle{Ideal Hyperbolic Polyhedra and Discrete Uniformization}.
\newblock \bibinfo{journal}{\emph{Discrete \& Computational Geometry}} \bibinfo{volume}{64}, \bibinfo{number}{1} (\bibinfo{year}{2020}), \bibinfo{pages}{63--108}.
\newblock


\bibitem[Springborn et~al\mbox{.}(2008)]%
        {springborn2008conformal}
\bibfield{author}{\bibinfo{person}{Boris Springborn}, \bibinfo{person}{Peter Schr{\"o}der}, {and} \bibinfo{person}{Ulrich Pinkall}.} \bibinfo{year}{2008}\natexlab{}.
\newblock \showarticletitle{Conformal Equivalence of Triangle Meshes}.
\newblock \bibinfo{journal}{\emph{ACM Transactions on Graphics}} \bibinfo{volume}{27}, \bibinfo{number}{3} (\bibinfo{year}{2008}), \bibinfo{pages}{1--11}.
\newblock


\bibitem[Tong et~al\mbox{.}(2006)]%
        {tong2006dqd}
\bibfield{author}{\bibinfo{person}{Y. Tong}, \bibinfo{person}{P. Alliez}, \bibinfo{person}{D. Cohen-Steiner}, {and} \bibinfo{person}{M. Desbrun}.} \bibinfo{year}{2006}\natexlab{}.
\newblock \showarticletitle{{Designing quadrangulations with discrete harmonic forms}}.
\newblock \bibinfo{journal}{\emph{Symposium on Geometry Processing}} (\bibinfo{year}{2006}), \bibinfo{pages}{201--210}.
\newblock


\bibitem[Vaxman et~al\mbox{.}(2016)]%
        {Vaxman:FieldsSTAR}
\bibfield{author}{\bibinfo{person}{Amir Vaxman}, \bibinfo{person}{Marcel Campen}, \bibinfo{person}{Olga Diamanti}, \bibinfo{person}{Daniele Panozzo}, \bibinfo{person}{David Bommes}, \bibinfo{person}{Klaus Hildebrandt}, {and} \bibinfo{person}{Mirela Ben-Chen}.} \bibinfo{year}{2016}\natexlab{}.
\newblock \showarticletitle{Directional Field Synthesis, Design, and Processing}.
\newblock \bibinfo{journal}{\emph{Comp. Graph. Forum}} \bibinfo{volume}{35}, \bibinfo{number}{2} (\bibinfo{year}{2016}).
\newblock


\bibitem[Zhou et~al\mbox{.}(2020)]%
        {zhou2020combinatorial}
\bibfield{author}{\bibinfo{person}{Jiaran Zhou}, \bibinfo{person}{Changhe Tu}, \bibinfo{person}{Denis Zorin}, {and} \bibinfo{person}{Marcel Campen}.} \bibinfo{year}{2020}\natexlab{}.
\newblock \showarticletitle{Combinatorial construction of seamless parameter domains}. In \bibinfo{booktitle}{\emph{Computer Graphics Forum}}, Vol.~\bibinfo{volume}{39}. Wiley Online Library, \bibinfo{pages}{179--190}.
\newblock


\bibitem[Zhou and Jacobson(2016)]%
        {Thingi10K}
\bibfield{author}{\bibinfo{person}{Qingnan Zhou} {and} \bibinfo{person}{Alec Jacobson}.} \bibinfo{year}{2016}\natexlab{}.
\newblock \showarticletitle{Thingi10K: A Dataset of 10,000 3D-Printing Models}.
\newblock \bibinfo{journal}{\emph{arXiv preprint arXiv:1605.04797}} (\bibinfo{year}{2016}).
\newblock


\end{thebibliography}
\newpage
\clearpage

\newpage
\clearpage
\end{document}